\documentstyle[aps,prb,eqsecnum,preprint]{revtex}

\input{epsf}

\begin{document}

\draft


\title{Many-Electron Trial Wave Functions for Inhomogeneous Solids}

\author{R. Gaudoin, M. Nekovee,\thanks{Present address: Centre for 
Computational Science, Department of Chemistry, Queen Mary and
Westfield College, Mile End Road, London E1 4NS, England} and
W. M. C. Foulkes}
\address{CMTH Group, Blackett Laboratory, Imperial College 
of Science, Technology and Medicine, Prince Consort Road, 
London SW7 2BZ, England}

\author{R. J. Needs and G. Rajagopal}
\address{TCM Group, Cavendish Laboratory, Cambridge University,
Madingley Road, Cambridge CB3 0HE}

\date{\today}

\maketitle

\begin{abstract}
Quantum Monte Carlo simulations of interacting electrons in solids
often use Slater-Jastrow trial wave functions with Jastrow factors
containing one- and two-body terms.  In uniform systems the long-range
behavior of the two-body term may be deduced from the random-phase
approximation (RPA) of Bohm and Pines.  Here we generalize the RPA to
nonuniform systems.  This gives the long-range behavior of the
inhomogeneous two-body correlation term and provides an accurate
analytic expression for the one-body term.  It also explains why
Slater-Jastrow trial wave functions incorporating determinants of
Hartree-Fock or density-functional orbitals are close to optimal even
in the presence of an RPA Jastrow factor.  After adjusting the
inhomogeneous RPA Jastrow factor to incorporate the known short-range
behavior, we test it using variational Monte Carlo calculations.  We
find that the most important aspect of the two-body term is the
short-range behavior due to electron-electron scattering, although the
long-range behavior described by the RPA should become more important
at high densities.
\end{abstract}



\section{Introduction} 
\label{sec:intro}

This paper discusses approximate ground-state wave functions for
inhomogeneous interacting many-electron systems such as solids.  In
particular, we consider wave functions of the Slater-Jastrow type,
$\Psi = e^{J} D$, where $D$ is a Slater determinant and $J$, the
Jastrow factor, takes account of the electronic correlations.  We have
two main aims: we wish to devise a method for generating inhomogeneous
Jastrow factors appropriate for use in strongly inhomogeneous solids;
and we wish to understand the surprisingly accurate results obtained
when Slater-Jastrow trial functions are used in variational (V) and
diffusion (D) quantum Monte Carlo (QMC)
simulations\cite{kalos_1986,hammond_1994} of weakly-correlated solids
such as silicon.  Despite the apparent simplicity of the
Slater-Jastrow form, cohesive energies calculated using VQMC are
typically an order of magnitude more
accurate\cite{fahy_1988,fahy_1990,li_1991,rajagopal_1995,kent_1999}
than cohesive energies obtained using Hartree-Fock (HF) or
density-functional theory within the local density approximation
(LDA).\cite{parr_1989} Cohesive energies calculated using DQMC are
even more accurate.

This is not the place for a general introduction to QMC (see Hammond
{\it et al.}\cite{hammond_1994} for a review), but a brief sketch of
the VQMC method may be helpful.  The idea is to obtain an approximate
many-electron ground state by numerically optimizing an explicit
parametrized trial wave function.  Once this has been done, the
calculation of expectation values reduces to the evaluation of
multi-dimensional integrals.  Ordinary integration methods using a
grid become very inefficient when the dimension of the integral is
greater than 5 or 10 (equivalent to a 2 or 3 electron system in three
dimensions), and so the integrals are evaluated using Monte Carlo
integration.\cite{metro_1953} This approach scales much more favorably
with the dimension than grid-based integration methods.

The trial wave functions used in most QMC simulations contain a Slater
determinant of LDA or HF orbitals and a Jastrow factor that includes
pairwise correlation terms, $u_{\sigma_i,\sigma_j}({\bf r}_i,{\bf
r}_j)$, and one-electron terms, $\chi_{\sigma_i}({\bf r}_i)$, where
${\bf r}_i$ and $\sigma_i$ are the position and spin component of
electron $i$.  The $u$ and $\chi$ functions are usually obtained by
optimizing specific parametrized functional forms according to the
variational principle.  In simulations of solids, it is common to
simplify the optimization by insisting that $u$ be both homogeneous
and isotropic (i.e., $u_{\sigma_i,\sigma_j}({\bf r}_i,{\bf r}_j)$ is
assumed to depend only on the interelectronic distance $r_{ij} = |{\bf
r}_i - {\bf r}_j|$).

By contrast, our principal aim is to derive a physically-motivated
inhomogeneous and anisotropic Jastrow factor for nonuniform systems,
based on a generalization of the random-phase approximation (RPA) of
Bohm and Pines.\cite{Bohm:RPA} This will enable us to reduce or even
dispense with the time-consuming optimization procedure.  The RPA is
known to give unphysical results in some cases and is not generally
regarded as an accurate quantitative method.  Here, however, we use
the RPA only to guide the construction of an approximate trial wave
function.  Once this wave function has been chosen, energies are
calculated using the true interacting many-electron Hamiltonian
instead of the approximate RPA form.  The results are therefore
variational and much more accurate than standard RPA energies.

The RPA theory of the \emph{homogeneous} electron gas is already used
in the construction of Jastrow factors for QMC
simulations.\cite{ceperley_1978} This approach predicts that the $u$
function should decay like $1/r_{ij}$ for large $r_{ij}$, but says
nothing about the $\chi$ function (which is zero in a homogeneous
system) or the short-range behavior of $u$.  Although it is easy to
modify the $u$ function to make it have the correct cusp-like behavior
at short range, the absence of $\chi$ terms implies that the
homogeneous RPA Jastrow factor produces inaccurate densities, and
hence poor energies, when used in strongly inhomogeneous systems.
This problem is usually fixed by adding a parametrized $\chi$ function
and optimizing the parameters numerically.\cite{fahy_1988}

Here, we generalize the RPA theory to \emph{inhomogeneous} systems
(see Malatesta\cite{malatesta_1997} and Fahy\cite{Fahy:CHI} for an
alternative approach).  We obtain a correlation term that is truly
inhomogeneous and anisotropic.  Furthermore, as originally noted by
Malatesta {\it et al.},\cite{malatesta_1997} we find that the
inhomogeneous RPA theory automatically produces a Jastrow factor
containing $\chi$ terms.  The short-range behavior of the
inhomogeneous RPA $u$ function is no better than that of the
homogeneous version but is more difficult to correct.  We impose the
required short-range cusps using a $k$-space method which, although
approximate, works well.

An interesting aspect of the inhomogeneous RPA theory is that it
provides a justification for using Slater determinants consisting of
LDA or HF orbitals.  This is common practice but is not obviously
correct: one might guess that the HF orbitals that are optimal in the
absence of a Jastrow factor would no longer be accurate in its
presence.  In fact, the RPA theory shows that HF or LDA orbitals are
close to optimal whether or not an RPA Jastrow factor is present.

We test our inhomogeneous RPA Jastrow factor by carrying out VQMC
calculations for a strongly inhomogeneous electron gas.  We find that
the inhomogeneous RPA $\chi$ functions are of such high quality that
there is no need to resort to the standard but costly numerical
optimization methods.  Surprisingly, however, we gain almost no
advantage by using inhomogeneous $u$ functions: a Jastrow factor
consisting of a homogeneous $u$ function plus any good $\chi$ function
gives almost the same results.  Inhomogeneous $u$ functions are often
used in full core atomic and molecular
calculations\cite{schmidt_1992,huang_1997} but do not seem necessary
in the strongly inhomogeneous electron gases studied here.  Some
recent VQMC simulations by Hood\cite{hood_1997,hood_1998} and
Nekovee\cite{nekovee_1999} suggest a possible explanation for this
observation.  Although our system has a strongly inhomogeneous
exchange-correlation hole $n_{xc}({\bf r},{\bf r}') = [ g({\bf r},{\bf
r}') - 1 ] n({\bf r}')$, the LDA seems to provide a better description
of the pair-correlation function $g({\bf r},{\bf r}')$.  The use of a
strongly inhomogeneous $u$ function may therefore be unnecessary.

In agreement with previous work\cite{malatesta_1997} we conclude that
the most important features of the Jastrow factor are (i) that the
corresponding Slater-Jastrow wave function produces an accurate
electron density, and (ii) that the $u$ function has the correct
cusp-like behavior whenever two electrons approach each other.  As
long as both these conditions are satisfied, the calculated energies
are normally quite accurate.

In summary, the five main results of our paper are:
\begin{enumerate}
\item
We have extended the RPA to inhomogeneous systems.
\item
We have used the inhomogeneous RPA theory to derive expressions for
the $\chi$ function and for the long-range behaviour of the fully
inhomogeneous $u$ function.
\item
We have developed a general method for imposing a cusp on any
inhomogeneous Jastrow factor expressed in $k$-space.
\item
The inhomogeneous RPA analysis has enabled us to explain why
Slater-Jastrow-type wave functions containing LDA or HF orbitals work
so well.
\item
We have implemented and tested the inhomogeneous RPA Jastrow factor
and found that the calculated one-body term works perfectly, but that
the inhomogeneity of the two-body term contributes little to the
energy at typical valence electron densities.
\end{enumerate}

The rest of this paper is organized as follows.  In Sec.~\ref{sec:QMC}
we describe the Slater-Jastrow trial wave functions used in most QMC
simulations of atoms, molecules, and solids.  Section \ref{sec:RPA}
presents the RPA theory of the inhomogeneous electron gas and explains
how it leads to Slater-Jastrow trial wave functions containing both
$\chi$ terms and inhomogeneous $u$ terms.  To supplement the somewhat
mathematical presentation in Sec.~\ref{sec:RPA}, Appendix
\ref{sec:appA} describes the RPA in more physical terms.  Section
\ref{sec:results} discusses the results of the VQMC simulations we
have done to test the inhomogeneous RPA Jastrow factor, and
Sec.~\ref{sec:conclusions} concludes.

\section{Trial wave functions for QMC simulations}
\label{sec:QMC}

The aim of this paper is to provide a better physical understanding of
the success of the Slater-Jastrow trial wave functions used in many
QMC calculations of atoms, molecules, and weakly correlated solids.  A
Slater-Jastrow trial function is the product of a totally
antisymmetric Slater determinant $D$ and a totally symmetric Jastrow
factor $e^{J}$.  The Slater determinant is often split into two
smaller determinants, one for each spin value:

\begin{equation}
\label{sjw}	
\Psi = e^{J} D_{\uparrow} D_{\downarrow} \;.
\end{equation} 

\noindent
This spoils the antisymmetry of the trial wave function on interchange
of electrons of opposite spin, but does not affect expectation values
of spin-independent operators.\cite{foulkes_1999} Since two smaller
determinants are easier to deal with than one big one, it also reduces
the numerical complexity of the problem.  The orbitals used in
$D_{\uparrow}$ and $D_{\downarrow}$ are normally obtained from LDA or
HF calculations.

The Slater determinants build in exchange effects but neglect the
electronic correlations caused by the Coulomb interactions.  The most
important correlation effects occur when pairs of electrons approach
each other, and these may be included by choosing a pairwise Jastrow
factor of the form,

\begin{equation}
J=\frac{1}{2} \sum_{i, j} 
u_{\sigma_i \sigma_j}({\bf r}_{i},{\bf r}_{j}) \;,
\end{equation} 

\noindent
where $u_{\sigma \sigma'}({\bf r},{\bf r}')=u_{\sigma' \sigma}({\bf
r}',{\bf r})$.  Because the LDA or HF orbitals in $D_{\uparrow}$ and
$D_{\downarrow}$ already give a reasonably good approximation to the
density, the introduction of a two-body $u$ function usually causes
the density of the many-electron wave function to deteriorate.  As a
result the trial energy deteriorates too.  It is therefore necessary
to introduce one-body $\chi$ terms to adjust the Jastrow factor:

\begin{equation}
\label{Jdef}
e^{J}=\exp \left( \frac{1}{2}
\sum_{i,j} u_{\sigma_i \sigma_j}({\bf r}_{i}, {\bf r}_{j}) + 
\sum_{i} \chi_{\sigma_i}({\bf r}_{i}) \right) \;.
\end{equation} 

Note the sign conventions here: our definition of $u$ is the negative
of that used by many other authors.  Most authors also omit the
diagonal $i=j$ terms in the sum over $i$ and $j$, but we find it
mathematically convenient to include them in our analysis.  In
homogeneous systems the diagonal terms only affect the normalization
of the trial function, while in inhomogeneous systems they add
one-body contributions that may be accounted for by a simple
redefinition of $\chi_{\sigma}({\bf r})$.

Since we are using periodic boundary conditions, it will often prove
convenient to express the electron density and Jastrow factor in
reciprocal space:

\begin{eqnarray}
n_{\sigma}({\bf k}) & = & \frac{1}{\sqrt{V}} \int_{V} 
n_{\sigma}({\bf r})e^{i{\bf k} 
\cdot {\bf r}} d^{3}r \;, \\
\chi_{\sigma}({\bf k}) & = & \frac{1}{\sqrt{V}} \int_{V} 
\chi_{\sigma}({\bf r})e^{i{\bf k}\cdot{\bf r}} d^{3}r \;, \\
u_{\sigma\sigma'}({\bf k},{\bf k}') & = & \frac{1}{V} \int_{V}
e^{i{\bf k}\cdot{\bf r}}
u_{\sigma\sigma'}({\bf r},{\bf r}')
e^{-i{\bf k}' \cdot {\bf r}'}d^{3}r d^{3}r' .
\end{eqnarray} 

\noindent
Note that we are using a symmetric definition of the Fourier
transformation; the corresponding back transformation is $f({\bf
r})=V^{-1/2} \sum_{\bf k} f({\bf k})e^{-i{\bf r}\cdot{\bf k}}$.

Several different types of Jastrow factor are in common use:

\begin{enumerate}

\item Jastrow factors based on the RPA theory of the uniform electron
gas in the form due to Bohm and Pines.\cite{Bohm:RPA} (In this paper
we show how to generalization this approach to obtain Jastrow factors
for strongly inhomogeneous systems.)

\item Jastrow factors based on the RPA theory of the uniform electron
gas in the form due to Gaskell.\cite{gaskell_1961} This type of
Jastrow factor has been widely used by Ceperley and
coworkers.\cite{ceperley_1978}

\item Jastrow factors obtained by the numerical optimization of a
parametrized functional form.  In most cases the optimization is
carried out using the variance minimization technique in the form
developed by Umrigar.\cite{umrigar_1988,PRCK}

\item Jastrow factors derived using the Fermi hypernetted 
chain approximation as given by Krotscheck et al.
\cite{FHNC1}

\end{enumerate}

\subsection{The cusp conditions}

In this section we deduce the short-range behavior of $u$.  When the
distance $r_{ij}=|{\bf r}_{i}-{\bf r}_{j}|$ between two electrons
approaches zero, the potential energy in the Hamiltonian operator
diverges like $1/r_{ij}$.  (Except where otherwise stated, we use
Hartree atomic units, $\hbar = e = 4\pi\varepsilon_0 = m_e = 1$,
throughout this paper.)  Since, for any exact eigenstate,
$\hat{H}\Psi$ is proportional to $\Psi$, this Coulomb divergence must
be cancelled by an equal and opposite divergence in the kinetic energy
terms.  When choosing a trial wave function for use in a QMC
simulation, it is important to ensure that this exact cancellation
still occurs.  If it does not, the local energy
$E_{L}={\hat{H}\Psi}/{\Psi}$, which is the quantity actually sampled
in the simulation, diverges as $r_{ij}$ approaches zero, causing
numerical difficulties.

Since, away from the nuclei, the single-particle orbitals are smooth
functions of the electronic coordinates, the determinantal part of the
trial wave function depends smoothly on the positions of the electrons.
For small ${\bf r}_{ij}$, it may therefore be expanded in the form:
\begin{equation}
\left. \left ( D_{\uparrow}D_{\downarrow} \right )
\right|_{r_{ij}\rightarrow 0}=c+{\bf a}\cdot{\bf r}_{i j}+\ldots \;.
\end{equation}
The application of the kinetic energy operator to this series produces
a smooth function of ${\bf r}_{ij}$, whereas the application of the
potential energy operator places a divergent $1/r_{ij}$ factor in
front of every term.

It is clear that a trial function containing only the Slater
determinants does not show the required cancellation of potential and
kinetic energy divergences.  The cancellation may however be imposed
by modifying the terms of the expansion as follows:
\begin{equation}
\begin{array}{lll}
c & \rightarrow & (1+\frac{1}{2}r_{i j})\,c \;, \\
{\bf a}\cdot{\bf r}_{i j} & \rightarrow & 
\rule{0mm}{4mm} (1+\frac{1}{4}r_{i j})\,{\bf a}\cdot{\bf r}_{i j} \;.
\end{array}  
\end{equation}
Each term in the series is multiplied by a factor, $1 + \alpha
r_{ij}$, which includes a cusp at $r_{ij} = 0$.  The numerical
constant $\alpha$ in front of the cusp is $1/(2 + 2n)$, where $n$ is
the order of ${\bf r}_{i j}$ in the original expansion.

A convenient way to introduce a cusp-like term is to include it in the
Jastrow factor $e^{J}$ in such a way that the cusp terms 
$(1 + \frac{1}{2} r_{ij})$ or $(1+\frac{1}{4}r_{i j})$ appear 
in the expansion of $e^{J}$ for small $r_{ij}$.  
As we only have one Jastrow factor we can
only hope to deal with one of the terms in the expansion of
$D_{\uparrow}D_{\downarrow}$ correctly; we choose the lowest order
term that is nonzero, as this is the one that causes the divergence in
the local energy $E_{L}$.  If electrons $i$ and $j$ have antiparallel
spins, the lowest order term is the constant $c$; for parallel spins,
$D_{\uparrow}D_{\downarrow}$ is an antisymmetric function of ${\bf
r}_{ij}$ and so the lowest order term is ${\bf a}\cdot{\bf r}_{i j}$.
The required cusp may thus be introduced by imposing the following
conditions on $u_{\sigma_i\sigma_j}$:
\begin{equation}
\label{cca}
\left. \frac{\partial u_{\uparrow\downarrow}({\bf r}_i,{\bf r}_j)}
{\partial r_{ij}} \right|_{r_{ij}=0}=\frac{1}{2}
\end{equation}

\noindent
for antiparallel spins, and

\begin{equation}
\label{ccp}
\left. \frac{\partial u_{\uparrow\uparrow}({\bf r}_i,{\bf r}_j)}
{\partial r_{ij}} \right|_{r_{ij}=0}=\frac{1}{4}
\end{equation}

\noindent
for parallel spins.\cite{kato_1957,pack_1966}

\subsection{The homogeneous RPA correlation term}
\label{subsec:QMC-RPAu}

As will be explained in Sec.~\ref{sec:RPA}, the RPA theory of Bohm and
Pines\cite{Bohm:RPA} suggests that the long-range behavior of the
correlation term $u$ in a homogeneous electron gas of number density
$n$ should take the form:

\begin{equation}
\label{BPf}
u_{\sigma_i\sigma_j}({\bf r}_{i},{\bf r}_{j}) = 
u_{\sigma_i \sigma_j}(r_{ij})= 
-\frac{1}{\omega_{p} r_{ij}} \;,
\end{equation}

\noindent
where $\omega_{p}=\sqrt{4\pi n}$ is the plasma frequency.  This
spin-independent two-body term has no cusp and is only expected to be
correct for large $r_{ij}$.  Multiplying Eq.~(\ref{BPf}) by
$1-e^{-r_{ij}/F_{\sigma_i \sigma_j}}$ yields

\begin{equation}
\label{homu}
u_{\sigma_i \sigma_j}(r_{ij}) =
-\frac{1}{\omega_{p} r_{ij}} \left (
1-e^{-r_{ij}/F_{\sigma_i \sigma_j}} \right ) \;,
\end{equation}

\noindent
which approaches Eq.~(\ref{BPf}) for large $r_{ij}$ and also has the
correct cusp behavior if $F_{\sigma_i \sigma_j}$ is chosen
appropriately.  Because of the spin dependence of the cusp conditions,
we need different values of $F$ for parallel- and antiparallel-spin
electrons.  By expanding Eq.~(\ref{homu}) to first order in $r_{ij}$,
we see that the cusp conditions Eq.~(\ref{cca}) and Eq.~(\ref{ccp})
become

\begin{equation}
\label{eq:f1}
\frac{1}{2}=
\left.\frac{\mbox{d}u_{\uparrow\downarrow}}{\mbox{d}r}\right|_{r=0}=
\frac{1}{2F_{\uparrow\downarrow}^{2}\omega_{p}}
\end{equation}

\noindent
for antiparallel spins, and

\begin{equation}
\label{eq:f2}
\frac{1}{4}=
\left.\frac{\mbox{d}u_{\uparrow\uparrow}}{\mbox{d}r}\right|_{r=0}=
\frac{1}{2F_{\uparrow\uparrow}^{2}\omega_{p}}
\end{equation}

\noindent
for parallel spins.

\subsection{The $\chi$ function}
\label{subsec:QMC-chi}

The $\chi$ function needed to counteract the density-adjusting effects
of the $u$ function is normally obtained by numerical optimization.
If we wish to avoid this costly procedure we have several options:

\begin{enumerate}

\item We could use no one-body term and hope for the best.  This is
the correct thing to do in a homogeneous system and also proves
satisfactory in cases when the spatial variation of the charge density
is much more rapid than that of the $u$ function.

\item We could use the result of Sec.~\ref{sec:RPA} and include a
one-body term of the form:
\begin{equation}
\label{fchi}
\chi_{\sigma}({\bf k})= -\sum_{{\bf k}',\sigma'} 
u_{\sigma\sigma'}({\bf k},{\bf k}') n_{\sigma'}({\bf k}') \;.
\end{equation}

\item We could use a one-body term as given by Malatesta 
{\it et al.}\cite{malatesta_1997}:
\begin{equation}
\label{fahychi}
\chi_{\sigma}({\bf k})= - \sqrt{V} 
\sum_{\sigma'} u_{\sigma\sigma'}({\bf k})
n_{\sigma'}({\bf k}) \;.
\end{equation}

\end{enumerate}

\noindent
In cases when the $u$ function is homogeneous,

\begin{equation}
\label{two1here}
u_{\sigma\sigma'}({\bf r},{\bf r}')=u_{\sigma\sigma'}({\bf r}-{\bf r}') \;,
\end{equation}

\noindent
we get

\begin{equation}
u_{\sigma\sigma'}({\bf k},{\bf k}') = 
\sqrt{V} u_{\sigma\sigma'}({\bf k}) \delta_{{\bf k},{\bf k}'}
\end{equation}

\noindent
and so Eq.~(\ref{fchi}) reduces to Eq.~(\ref{fahychi}).  Option (2)
therefore reduces to option (3).  Strictly speaking, however, $u$ is
never exactly homogeneous unless the electron density is exactly
uniform, in which case both options (2) and (3) merely state that
$\chi = 0$.

\section{The random-phase approximation for inhomogeneous systems}
\label{sec:RPA}

This section starts with a brief introduction to the ideas behind the
standard RPA treatment of homogeneous
systems.\cite{Bohm:RPA,Bohm:SUBC,Raimes:MET} We then generalize the
RPA theory to inhomogeneous systems.  This generalization shows us how
to construct approximate ground-state wave functions of the
Slater-Jastrow type.  The RPA forms of the $u$ and $\chi$ functions
are closely related and both depend on the electron density.

\subsection{Review of the RPA for homogeneous systems}
\label{subsec:RPAreview}

Let us first give a brief overview of the standard RPA as formulated
for homogeneous systems.  The starting point is the usual Hamiltonian,

\begin{equation}
\label{eq:hbasic}
\hat{H} = \frac{1}{2}\sum_{i} \hat{\bf p}_{i}^{2}    
+ 2\pi  \sum_{\bf k} 
\frac{\hat{n}_{\bf k}\hat{n}^{\dagger}_{\bf k}}{k^2}
-\frac{2\pi N}{V}\sum_{\bf k} \frac{1}{k^{2}} \;,
\end{equation}

\noindent
of a uniform electron gas in a volume $V$.  Note that we have adopted
a slightly unusual definition of the number density operator,

\begin{equation}
\label{eq:nk}
\hat{n}_{\bf k}=\frac{1}{\sqrt{V}}\sum_{i}e^{i {\bf k} \cdot 
\hat{\bf r}_{i}} \;,
\end{equation}

\noindent
which includes a $1/\sqrt{V}$ factor in order to be consistent with
the symmetric definition of the Fourier transform used throughout this
paper.  The $k$-points are chosen in accordance with the geometry of
the system, which is taken to obey periodic boundary conditions.  The
system is assumed to be charge neutral and so the ${\bf k}=0$ terms
are omitted from the $k$-space summations.

As we are interested in describing the long-range correlations due to
the electron-electron interaction, we wish to make a connection
between the Hamiltonian of Eq.~(\ref{eq:hbasic}) and the long
wavelength density fluctuations known as plasmons.  Our eventual aim
is to split the Hamiltonian into an electronic term with short-range
interactions and a plasmon term that is only weakly coupled to the
electrons.  Once the electron and plasmon parts of the Hamiltonian
have been (almost) decoupled in this way, we shall see that the
long-range correlations are described by the ground-state wave function
of the plasmon part, which is simply a set of quantum mechanical
harmonic oscillators.  At shorter wavelengths the plasmons are not
well defined and the collective plasmon description of electronic
correlation ceases to be valid.  Instead, the electrons feel a
short-range screened interaction that produces additional
electron-electron scattering-like correlations.

The assumption of weak plasmon-electron interaction is reasonable at
small $k$ since long wavelength plasmons are long lived.  For larger
$k$ values, however, the almost flat plasmon dispersion curve runs
into the continuum of electron-hole pair excitations and the plasmons
are no longer well defined.  In a uniform electron gas this
happens at a wave vector $k_c$ given by\cite{Raimes:MET}

\begin{equation}
\label{cutoffk}
k_{c} \approx \frac{1}{2}k_{F} r_{s}^{{1}/{2}} \propto 
n^{{1}/{6}} \;,
\end{equation}

\noindent
where $k_F$ is the Fermi wave vector, $r_s a_0$ is the radius of a
sphere containing one electron on average, and $a_0$, the Bohr radius,
is the atomic unit of length.  This estimate of the cutoff should also
be applicable to inhomogeneous systems as long as the density does not
vary too much.  We see that for typical metals with $r_s$ values of 2 or
3 the cutoff is of the order of the Fermi wave vector.  We also see that
the higher the density the better the plasmon description should be. The
inverse of $k_c$ is a measure of the minimum length scale on which the
electronic correlations may be described in terms of plasmons.

The first step in the derivation of the homogeneous RPA\cite{Bohm:RPA}
is to introduce a new pair of conjugate operators, $\hat{\pi}^{ }_{\bf
k}$ and $\hat{q}^{ }_{\bf k}$, for every wave vector ${\bf k}$ with
modulus $k < k_{c}$.  These operators act in a new Hilbert space that
we call the \emph{oscillator space}, and transform like $\hat{n}_{\bf
k}$ under Hermitian conjugation: $\hat{\pi}^{ }_{\bf k} =
\hat{\pi}^{\dagger}_{-{\bf k}}$ and $\hat{q}^{ }_{\bf k} =
\hat{q}^{\dagger}_{-{\bf k}}$.  The physical and oscillator Hilbert
spaces are quite distinct, and so $\hat{\pi}^{ }_{\bf k}$ and
$\hat{q}^{ }_{\bf k}$ commute with $\hat{\bf r}_{i}$ and $\hat{\bf
p}_{i}$.  For the time being $\hat{\pi}^{ }_{\bf k}$ and $\hat{q}^{
}_{\bf k}$ have little physical meaning, but later in the derivation
they will become associated with the plasmon coordinates.

The oscillator-space operators are now used to define a new
Hamiltonian,

\begin{eqnarray}
\label{newH}
\hat{H}_{\rm BP} & = & \frac{1}{2} \sum_{i} \hat{\bf p}_{i}^{2} + 
2\pi  \sum_{\bf k}  
\frac{\hat{n}^{ }_{\bf k}\hat{n}^{\dagger}_{\bf k}}{k^2}
-\frac{2\pi N}{V} \sum_{\bf k} \frac{1}{k^{2}}\nonumber \\
& & +\frac{1}{2}\sum_{k<k_{c}}\hat{\pi}^{ }_{\bf k}
\hat{\pi}^{\dagger}_{\bf k}
-\sum_{k<k_{c}} \left(\frac{4\pi }{k^{2}}\right)^{1/2}
\hat{\pi}^{ }_{\bf k}\hat{n}^{\dagger}_{\bf k} \;,
\end{eqnarray}

\noindent
which acts in an enlarged Hilbert space that is the product of the
physical space and the oscillator space.  The Bohm-Pines Hamiltonian
$\hat{H}_{\rm BP}$ may be written in the form

\begin{eqnarray}
\lefteqn { \hat{H}_{\rm BP} = \frac{1}{2} \sum_{i} \hat{\bf p}_{i}^{2} + 
2\pi  \sum_{k>k_{c}}  
\frac{\hat{n}^{ }_{\bf k}\hat{n}^{\dagger}_{\bf k}}{k^2}
-\frac{2\pi N}{V} \sum_{\bf k} \frac{1}{k^{2}} } \nonumber \\
& & +\frac{1}{2}\sum_{k<k_{c}}\left(\hat{\pi}^{ }_{\bf k}
-\left(\frac{4\pi }{k^{2}}\right)^{1/2}
\hspace*{-2mm}\hat{n}^{ }_{\bf k}\right)
\left(\hat{\pi}^{ }_{\bf k}
-\left(\frac{4\pi }{k^{2}}\right)^{1/2}
\hspace*{-2mm}\hat{n}_{\bf k}\right)^{\dagger} \hspace*{-1mm},
\end{eqnarray}

\noindent
and so its eigenvalue spectrum is bounded below.

We now restrict our attention to those states $|\Phi\rangle$ from the
enlarged Hilbert space that satisfy the \emph{subsidiary condition}:

\begin{equation}
\label{subcon1}
\hat{\pi}_{\bf k}|\Phi\rangle=0 \;\;\; {\rm for\;\,all\;\,} k<k_c.
\end{equation}

\noindent
Any state obeying Eq.~(\ref{subcon1}) may be written in the form
$|\Phi\rangle = |\psi \rangle | {\mbox{\boldmath $\pi$}}$$=$${\bf 0}
\rangle$, where $|\psi\rangle$ is a state in the physical Hilbert
space and $| {\mbox{\boldmath $\pi$}}$$=$${\bf 0} \rangle$ is the
oscillator-space state satisfying

\begin{equation}
\hat{\pi}^{ }_{\bf k} | {\mbox{\boldmath $\pi$}}$$=$${\bf 0} \rangle =
0 \;\;\; {\rm for\;\,all\;\,}k<k_c.
\end{equation}

\noindent
(The bold symbol ${\mbox{\boldmath $\pi$}}$ is shorthand for the
vector of all the $\pi_{\bf k}$ with $k<k_c$.)  The set of
product-space states satisfying the subsidiary condition may therefore
be put into one-to-one correspondence with the set of physical states.
Equally, given any eigenstate $|\psi\rangle$ with eigenvalue $E$ of
the original Hamiltonian, the product state $|\Phi\rangle =
|\psi\rangle|{\mbox{\boldmath $\pi$}}$$=$${\bf 0}\rangle$ satisfies
the subsidiary condition, Eq.~(\ref{subcon1}), and is an eigenstate of
Eq.~(\ref{newH}) with the same energy.  This is because all the new
terms in $\hat{H}_{\rm BP}$ involve $\hat{\pi}^{ }_{\bf k}$ and hence
give zero when operating on $| {\mbox{\boldmath $\pi$}}$$=$${\bf0}
\rangle$.  As long as the subsidiary condition is obeyed, the
additional degrees of freedom may simply be regarded as dummy
variables, both in the wave function and the Hamiltonian.

It will be proved later on in this paper that the overall ground state
$|\Phi_{0}\rangle$ of the extended Hamiltonian, Eq.~(\ref{newH}),
\emph{automatically} satisfies the subsidiary condition,
Eq.~(\ref{subcon1}).  The physical-space part of the ground state of
$\hat{H}_{\rm BP}$ is therefore the ground state of the original
Hamiltonian, and we might as well study the ground-state properties of
the extended Hamiltonian as those of the original.

The next stage in the derivation is to make a unitary transformation,
which will be described in more detail in Sec.~\ref{alg.2}.  Once the
extended Bohm-Pines Hamiltonian has been transformed and several
supposedly small terms have been dropped, the transformed Hamiltonian
may be written as the sum of a spatial Hamiltonian with a short-range
interaction,

\begin{equation}
\label{srH}
\hat{H}_{sr} = \frac{1}{2}\sum_{i} \hat{\bf p}_{i}^{2} 
+2\pi\sum_{k>k_{c}} 
\frac{\hat{n}^{ }_{\bf k}\hat{n}^{\dagger}_{\bf k}}{k^2}
-\frac{2\pi N}{V} \sum_{\bf k} \frac{1}{k^{2}} \;,
\end{equation}

\noindent 
and an oscillator-like plasmon term,

\begin{equation}
\label{Hp}
\hat{H}_{p} = \frac{1}{2} \sum_{k<k_{c}}
\left(\hat{\pi}^{ }_{\bf k}\hat{\pi}^{\dagger}_{\bf k}
+ \omega_{p}^{2}\hat{q}^{ }_{\bf k}
\hat{q}^{\dagger}_{\bf k}\right) \;,
\end{equation}

\noindent
where $\omega_{p}=\left(4\pi N/V \right)^{1/2}$ is the plasma
frequency.  One of the approximations made during the derivation of
these results involves the replacement of an exponential factor of the
form $\exp[{i({\bf k}-{\bf k}') \cdot \hat{\bf r}_{i}}]$ by its
average value.  Since the electronic positions ${\bf r}_{i}$ are
effectively random in a homogeneous system, the phase of the
exponential is also random unless ${\bf k} = {\bf k}'$, and so the
average is $1$ if ${\bf k}={\bf k}'$ or zero otherwise.  This is the
eponymous random-phase approximation.

The approximate Hamiltonian,

\begin{equation}
\hat{H}_{\rm RPA}=\hat{H}_{sr}+\hat{H}_{p} \;,
\end{equation}

\noindent
still includes short-range electron-electron interactions and so
cannot be solved exactly.  We may, however, use the HF method or the
LDA to obtain approximate eigenstates of $\hat{H}_{sr}$.  These
approximate eigenstates, which are of course Slater determinants, may
then be multiplied by eigenstates of the oscillator Hamiltonian,
Eq.~(\ref{Hp}), to obtain approximate eigenstates of the full
Hamiltonian $\hat{H}_{\rm RPA}$.  The unitary transformation may then
be inverted to obtain approximate eigenstates of $\hat{H}_{\rm BP}$.
Unlike the exact ground state, the resulting approximate ground state
of $\hat{H}_{\rm BP}$ does not satisfy the subsidiary condition,
Eq.~(\ref{subcon1}), exactly.  It may, however, be projected onto the
subspace of states that do satisfy the subsidiary condition to obtain
an approximate ground state in the physical Hilbert space.  This
approximate physical ground state has the homogeneous Slater-Jastrow
form,

\begin{equation}
\psi_{0}(\{ {\bf x}_{i} \}) =
\exp\left[
-\frac{2\pi}{\omega_{p}}
\sum_{k<k_{c}} 
\frac{n^{ }_{\bf k}n^{\dagger}_{\bf k}}{k^2}
\right]  D(\{ {\bf x}_{i} \}) \;,
\end{equation} 

\noindent
where $n_{\bf k}=V^{-1/2}\sum_{i}e^{i{\bf k}\cdot{\bf r}_{i}}$ and
${\bf x}_i = ({\bf r}_i,\sigma_i)$ describes the position and spin
component of electron $i$.  The RPA $u$ function is therefore
independent of spin and proportional to $1/r_{ij}$ at large $r_{ij}$.
As noted earlier, the cutoff wave vector $k_{c}$ is usually chosen to
be of the order of the Fermi wave vector $k_{F}$.

\subsection{The RPA Hamiltonian in inhomogeneous systems}
\label{alg.1}

The aim is to find an approximate ground state $|\psi_{0}\rangle$ with
energy $E_{0}$ of the following Hamiltonian:

\begin{eqnarray}
\label{horig}
\hat{H}& =&
 \frac{1}{2}\sum_{i} \hat{\bf p}_{i}^{2}  + 
\sum_{i} V( \hat{\bf r}_{i})  
\nonumber \\
&+& 2\pi  \sum_{\bf k} 
\frac{ \hat{n}^{ }_{\bf k}\hat{n}^{\dagger}_{\bf k}}{k^2}
-\frac{2\pi N}{V}\sum_{\bf k} \frac{1}{k^{2}} \;,
\end{eqnarray}

\noindent
where $V({\bf r})$ is an applied potential.  The ``physical'' Hilbert
space in which this Hamiltonian acts will be denoted ${\cal H}_{R}$.
For reasons that will become clear later on, it is sometimes
convenient to rewrite $\hat{H}$ in the form,

\begin{eqnarray}
\hat{H}& =& \frac{1}{2} \sum_{i} \hat{\bf p}_{i}^{2} + 
\sum_{i} \tilde{V}( \hat{\bf r}_{i})  
\nonumber \\
&+& 2\pi  \sum_{\bf k} 
\frac{ \hat{n}^{ }_{\bf k}\hat{n}^{\dagger}_{\bf k}}{k^2}
-\frac{2\pi N}{V} \sum_{\bf k} \frac{1}{k^{2}}    \nonumber \\
&+& \frac{1}{2}\sum_{k<k_{c}}\pi^{0}_{\bf k}\pi^{0}_{-{\bf k}}
-\sum_{k<k_{c}} \left(\frac{4\pi }{k^{2}}\right)^{1/2}
\pi^{0}_{\bf k}\hat{n}^{\dagger}_{\bf k} \;,
\nonumber \\
\end{eqnarray}

\noindent
where $\tilde{V}({\bf r})$ is defined via:

\begin{eqnarray}
\label{npot}
&&\tilde{V}({\bf r})= V({\bf r}) -\frac{\Delta E}{N} - \Delta V({\bf
r}) \; , \\ 
&&\Delta E = \frac{1}{2}\sum_{k<k_{c}}\pi^{0}_{\bf k}
\pi^{0}_{-{\bf k}} \; , \\
&&\Delta V({\bf r}) =
-\sum_{k<k_{c}} \left(\frac{4\pi}{k^{2}}\right)^{1/2}
\pi^{0}_{\bf k} \, \frac{e^{-i{\bf k}\cdot {\bf r}}}{\sqrt{V}} \;,
\end{eqnarray}

\noindent
and the $\pi^{0}_{\bf k}$ are arbitrary numbers satisfying
$\left.\pi^{0}_{\bf k}\right.^{\ast}=\pi^{0}_{\bf -k}$.

Let us now introduce conjugate pairs of operators, $\hat{\pi}_{\bf k}$
and $\hat{q}_{\bf k}$, acting in a different Hilbert space ${\cal
H}_{O}$, which we call the oscillator space.  The ``momentum''
operator $\hat{\pi}_{\bf k}$ and the ``position'' operator
$\hat{q}_{\bf k}$ are taken to be the Fourier transforms of field
operators $\hat{\pi}({\bf r}) = \hat{\pi}^{\dagger}({\bf r})$ and
$\hat{q}({\bf r})=\hat{q}^{\dagger}({\bf r})$:

\begin{eqnarray}
\label{opdef1}
\hat{q}^{\dagger}_{\bf k}
&=&\frac{1}{\sqrt{V}}\int_{V} e^{i{\bf k}\cdot{\bf r}}
\hat{q}^{\dagger}({\bf r}) d^{3}r \;,
\end{eqnarray}
\begin{equation}
\label{opdef2}
\hat{\pi}_{\bf k}=\frac{1}{\sqrt{V}}\int_{V} 
e^{i{\bf k} \cdot {\bf r}}\hat{\pi}({\bf r}) d^{3}r \;,
\end{equation}

\noindent
which satisfy the commutation relation

\begin{equation}
\left[\hat{\pi}({\bf r}), \hat{q}({\bf r}') \right] = 
-i\delta({\bf r}-{\bf r}') \;.
\end{equation}

\noindent
It therefore follows that $\hat{\pi}^{\dagger}_{\bf k}=\hat{\pi}^{
}_{-{\bf k}}$ and $\hat{q}^{\dagger}_{\bf k} = \hat{q}^{ }_{-{\bf k}}$
obey

%
%

\begin{equation}
\left[\hat{\pi}_{\bf k},\hat{q}_{{\bf k}'}\right]=
-i\delta_{{\bf k},{\bf k}'} \;.
\end{equation}

We can form an extended Hilbert space by taking the product space
${\cal H}_{\rm ext}={\cal H}_{R}\bigotimes{\cal H}_{O}$.  If we denote
the identity operators in the real and oscillator spaces by
$\hat{1}_{R}$ and $\hat{1}_{O}$ respectively, we can define
extended-space operators such as $\hat{r}_{\rm ext} = \hat{r}_{R}
\otimes \hat{1}_{O}$ and $\hat{q}_{\rm ext} = \hat{1}_{R} \otimes
\hat{q}_{O}$ corresponding to any operator belonging to one or other
of the constituent Hilbert spaces.  It is obvious that all the
extended ``$R$'' operators commute with all the extended ``$O$''
operators.  From now on we shall omit the identity operators and drop
all ``$O$'', ``$R$'', and ``${\rm ext}$'' subscripts.  Operators will
always be denoted using hats, and so anything without a hat may be
assumed to be a (possibly complex) number.

Now consider the Hermitian \emph{extended Hamiltonian},

\begin{eqnarray}
\label{eq:hbp}
\hat{H}_{\rm BP}&=& \frac{1}{2} \sum_{i} \hat{\bf p}_{i}^{2} + 
\sum_{i} \tilde{V}( \hat{\bf r}_{i}) \nonumber
\\
&+&2\pi\sum_{\bf k}  
\frac{\hat{n}^{ }_{\bf k}\hat{n}^{\dagger}_{\bf k}}{k^2}
-\frac{2\pi N}{k^2} \sum_{\bf k} \frac{1}{k^{2}}\nonumber \\
&+& \frac{1}{2}\sum_{k<k_{c}}\hat{\pi}_{\bf k}\hat{\pi}_{-{\bf k}}
-\sum_{k<k_{c}} \left(\frac{4\pi }{k^{2}}\right)^{1/2}
\hat{\pi}^{ }_{\bf k}\hat{n}^{\dagger}_{\bf k} 
\;,
\end{eqnarray}

\noindent
which is an operator in the extended Hilbert space ${\cal H}_{\rm
ext}$.  Note that the potential $\tilde{V}$ is still the same as given
by Eq.~(\ref{npot}); the variables $\pi^{0}_{\bf k}$ that were used to
construct $\tilde{V}$ have not been replaced by operators in ${\cal
H}_{O}$, but remain ordinary complex numbers.  Because $\hat{H}_{\rm
BP}$ and $\hat{\mbox{\boldmath $\pi$}}$ commute they may be
diagonalized simultaneously, and hence all eigenstates of the extended
Hamiltonian $\hat{H}_{\rm BP}$ may be written in the form
$|\psi_{\mbox{\boldmath $\pi$}}\rangle | {\mbox{\boldmath
$\pi$}}\rangle$, where $\hat{\pi}_{\bf k}|{\mbox{\boldmath
$\pi$}}\rangle=\pi_{\bf k}|{\mbox{\boldmath $\pi$}}\rangle$ for all
$k<k_c$.

Let $|\psi_{0}\rangle$ and $E_0$ be the ground-state wave function and
ground-state eigenvalue of the physical Hamiltonian $\hat{H}$.  If we
define an eigenstate $|{\mbox{\boldmath $\pi$}}^{0}\rangle$ of the
$\hat{\pi}_{\bf k}$ operators such that $\hat{\pi}_{\bf
k}|{\mbox{\boldmath $\pi$}^{0}}\rangle = \pi^{0}_{\bf k} |
{\mbox{\boldmath $\pi$}^{0}}\rangle$ for all $k<k_c$, it follows that
$|\psi_{0}\rangle|{\mbox{\boldmath $\pi$}^{0}}\rangle$ is an
eigenstate of $\hat{H}_{\rm BP}$ with the same eigenvalue $E_{0}$.  It
need not, however, be the ground state of $\hat{H}_{\rm BP}$, which
might correspond to a different wave function $|\psi_{\rm
min}\rangle|{\mbox{\boldmath $\pi$}^{\rm min}}\rangle$.  The question
that now arises is how to choose the constants $\pi^{0}_{\bf k}$ (and
hence the modified potential $\tilde{V}$) such that
$|\psi_{0}\rangle|{\mbox{\boldmath $\pi$}^{0}}\rangle$ is in fact the
ground state of $\hat{H}_{\rm BP}$ and not just some other eigenstate.
In other words, we have to find a link between the $\pi^{0}_{\bf k}$
and $|\psi_{0}\rangle$.  This can be achieved using the
Hellmann-Feynman theorem.

Consider the lowest energy state $|\Phi_{\mbox{\scriptsize\boldmath
$\pi$}}\rangle = |\psi_{\mbox{\scriptsize\boldmath $\pi$}}\rangle |
{\mbox{\boldmath $\pi$}}\rangle$ corresponding to some fixed
oscillator-space eigenstate $|{\mbox{\boldmath $\pi$}}\rangle$.  The
energy eigenvalue of $|\Phi_{\mbox{\scriptsize\boldmath
$\pi$}}\rangle$ will be denoted $E^{\mbox{\scriptsize\boldmath
$\pi$}}$.  When $\hat{H}_{\rm BP}$ acts on
$|\Phi_{\mbox{\scriptsize\boldmath $\pi$}}\rangle$, the operators
$\hat{\pi}_{\bf k}$ may be replaced by their eigenvalues $\pi_{\bf
k}$, and hence $|\psi_{\mbox{\scriptsize\boldmath $\pi$}}\rangle$ is
in fact the ground state of

\begin{eqnarray}
\label{hpi}
\hat{H}^{\mbox{\scriptsize\boldmath $\pi$}}
& =& \frac{1}{2} \sum_{i}\hat{\bf p}_{i}^{2} + 
\sum_{i} \tilde{V}( \hat{\bf r}_{i})  \nonumber
\\
&+& 2\pi\sum_{\bf k} 
 \frac{\hat{n}^{ }_{\bf k}\hat{n}^{\dagger}_{\bf k}}{k^{2}}
-\frac{2\pi N}{V} \sum_{\bf k} \frac{1}{k^{2}}\nonumber \\
&+& \frac{1}{2}\sum_{k<k_{c}}\pi_{\bf k}\pi_{-{\bf k}}
-\sum_{k<k_{c}} \left(\frac{4\pi }{k^{2}}\right)^{1/2}
\pi^{ }_{\bf k}\hat{n}^{\dagger}_{\bf k} \;.
\end{eqnarray}

\noindent
Note that $\hat{H}^{\mbox{\scriptsize\boldmath $\pi$}}$ operates not
in the extended Hilbert space but in ${\cal H}_{R}$.  Furthermore, it
is important to keep in mind that $\tilde{V}$ is formed using the as
yet unknown numbers $\pi^{0}_{\bf k}$.  The overall ground state of
$\hat{H}_{\rm BP}$ corresponds to the minimum of
$E^{\mbox{\scriptsize\boldmath $\pi$}}$ with respect to
\mbox{\boldmath $\pi$}. As we are now looking at a standard quantum
mechanical problem formulated in the physical Hilbert space ${\cal
H}_{R}$, the Hellmann-Feynman theorem gives:

\begin{eqnarray}
\frac{\partial E^{\mbox{\scriptsize\boldmath $\pi$}}}{\partial\pi_{\bf k}}  & = & 
{\left\langle \psi_{\mbox{\scriptsize\boldmath $\pi$}} \left| \frac{\partial 
\hat{H}^{\mbox{\scriptsize\boldmath $\pi$}}}{\partial \pi_{\bf k}}
\right|\psi_{\mbox{\scriptsize\boldmath $\pi$}}\right\rangle} \\
\label{stationary}
& = & \pi_{-\bf k}-\left(\frac{4\pi }{k^{2}}\right)^{1/2}
\langle \psi_{\mbox{\scriptsize\boldmath $\pi$}} | 
\hat{n}^{\dagger}_{\bf k}|\psi_{\mbox{\scriptsize\boldmath $\pi$}}
\rangle \;,
\end{eqnarray}

\noindent
where $|\psi_{\mbox{\scriptsize\boldmath $\pi$}}\rangle$ is assumed
normalized.  If the overall ground state of $\hat{H}_{\rm BP}$ occurs
when $\mbox{\boldmath $\pi$} = \mbox{\boldmath $\pi$}^{\rm min}$, it
follows that:
 
\begin{equation}
\label{min1}
0=
\pi^{\rm min}_{-{\bf k}}-\left(\frac{4\pi}{k^{2}}\right)^{1/2}
\langle \psi_{\mbox{\scriptsize\boldmath $\pi$}^{\rm min}}| 
\hat{n}^{\dagger}_{{\bf k}} | 
\psi_{\mbox{\scriptsize\boldmath $\pi$}^{\rm min}}\rangle
\;. 
\end{equation}

We wish to choose the constants $\pi^{0}_{\bf k}$ such that
$\pi^{0}_{\bf k} = \pi^{\min}_{\bf k}$, since in this case we have
already argued that the physical-space part
$|\psi_{{\mbox{\scriptsize\boldmath $\pi$}^{0}}}\rangle$ of the
Bohm-Pines ground state

\begin{equation}
\label{gsa}
|\Phi_{{\mbox{\scriptsize\boldmath $\pi$}^{0}}}\rangle 
=
|\psi_{{\mbox{\scriptsize\boldmath $\pi$}^{0}}}\rangle 
|{\mbox{\boldmath $\pi$}}^{0}\rangle
\end{equation}

\noindent
is equal to the exact physical ground state $|\psi_0\rangle$.  The
sought after link between $|\psi_{0}\rangle$ and the $\pi^{0}_{\bf k}$
is therefore:

\begin{eqnarray}
\label{numb1}
\pi^{0}_{-{\bf k}}&=&
\left(\frac{4\pi }{k^{2}}\right)^{1/2}
\langle \psi_{0}|\hat{n}^{\dagger}_{{\bf k}} 
|\psi_{0}\rangle 
\nonumber \\
&=&
\left(\frac{4\pi }{k^{2}}\right)^{1/2}
\langle \hat{n}_{-{\bf k}}\rangle_0 \;,
\end{eqnarray}

\noindent
where $\langle \hat{n}_{\bf k}\rangle_0 = \langle\psi_{0}|\hat{n}_{\bf
k}|\psi_{0}\rangle$ is a Fourier component of the ground-state
electron density $n({\bf r})$.  Note that Eq.~(\ref{numb1}) is
\emph{not} an operator equation; it simply links one number,
$\pi_{-{\bf k}}^{0}$, to another, $\langle\hat{n}_{-{\bf
k}}\rangle_0$.  Since all the $\hat{\pi}_{\bf k}$ operators have
$k<k_c$, Eq.~(\ref{numb1}) only applies when $k<k_c$.

We still have to verify that the stationary point $\mbox{\boldmath
${\pi}$}^{\rm min}$ is in fact the minimum of
$E^{\mbox{\boldmath\scriptsize $\pi$}}$.  Eq.~(\ref{stationary})
determines the slope at any given ${\mbox{\boldmath $\pi$}}$.  The
matrix of second derivatives of $E^{\mbox{\scriptsize\boldmath
$\pi$}}$, which determines the curvature, may therefore be obtained by
differentiating Eq.~(\ref{stationary}):

\begin{eqnarray}
\frac{\partial}{\partial\pi^{ }_{-{\bf k}'}}
\frac{\partial}{\partial\pi^{ }_{\bf k}}
E^{\mbox{\scriptsize\boldmath $\pi$}} & = & 
\delta_{-{\bf k},-{\bf k}'} + \frac{4\pi }{kk'} 
\chi(-{\bf k},-{\bf k}') \\
& = & k \epsilon^{-1}(-{\bf k},-{\bf k}') / k' \;,
\end{eqnarray}

\noindent
where

\begin{equation}
\chi({\bf k},{\bf k}') 
= - \frac{k'}{\sqrt{4\pi}} 
\frac{\partial n_{\bf k}}{\partial \pi_{{\bf k}'}}
\end{equation}

\noindent
is the static susceptibility matrix and $\epsilon^{-1}({\bf k},{\bf
k}')$ is the inverse dielectric matrix.  If we make the reasonable
assumption that all the eigenvalues of the dielectric matrix are
positive,\cite{Pines:TQL,Pines:ELEX,Keldysh:DFCS} it follows that all
the eigenvalues of the matrix of second derivatives of
$E^{\mbox{\scriptsize\boldmath $\pi$}}$ are also positive.  The
stationary point given by Eq.~(\ref{min1}) is then a minimum.

Eq.~(\ref{min1}) is the generalization to inhomogeneous systems of the
``subsidiary condition'' of Bohm and Pines.\cite{Bohm:RPA} It can be
rewritten as

\begin{equation}
\label{sc1}
\hat{\Omega}_{\bf k}|\Phi\rangle=0 \;\;\;\;\;\;\; (k<k_c),
\end{equation}

\noindent
where

\begin{equation}
\hat{\Omega}_{\bf k}=
\hat{\pi}_{\bf k} - 
\left(\frac{4\pi }{k^{2}}\right)^{1/2}
\langle\hat{n}_{\bf k}\rangle_0 \hat{1} \;,
\end{equation}

\noindent
and has to be obeyed by $|\Phi_{0}\rangle$, the exact ground state of
$\hat{H}_{\rm BP}$.  In the case of a homogeneous system,
Eq.~(\ref{sc1}) reduces to $\hat{\pi}_{\bf k}|\Phi\rangle=0$ as
derived by Bohm and Pines.\cite{Bohm:RPA}

The effect of the subsidiary condition is to reduce the number of
degrees of freedom of the extended Hilbert space ${\cal H}_{\rm ext}$.
The reduced Hilbert space satisfying the subsidiary condition is
spanned by the set of eigenstates of $\hat{H}_{\rm BP}$ of the form
$|\psi\rangle |{\mbox{\boldmath $\pi$}}\rangle$, where

\begin{equation}
{\pi}_{\bf k} = \left(\frac{4\pi }{k^{2}}\right)^{1/2}
\langle\hat{n}_{\bf k}\rangle_0  \;\;\; {\rm for\;\,all\;\,} k<k_c.
\end{equation}

\noindent
If we have set the parameters $\pi^0_{\bf k}$ using Eq.~(\ref{numb1}),
the states $|\psi\rangle$ are then eigenfunctions of the
\emph{original} Hamiltonian, Eq.~(\ref{horig}).  This follows from the
definition of the potential $\tilde{V}({\bf r})$, which is such that
the Hamiltonian in Eq.~(\ref{hpi}) is the same as the original
Hamiltonian of Eq.~(\ref{horig}) when the value of $\mbox{\boldmath
$\pi$}$ is consistent with the subsidiary condition:
$\hat{H}^{\mbox{\scriptsize\boldmath $\pi$}} \left . \rule{0mm}{2.5mm}
\right |_{\pi_{\bf k}=\pi^{0}_{\bf k}}=\hat{H} $.  The subspace
singled out by the subsidiary condition Eq.~(\ref{sc1}) is thus
equivalent to the original Hilbert space ${\cal H}_{R}$.

Let us now use the subsidiary condition to evaluate $\Delta E$ and
$\Delta V({\bf r})$.  For $\Delta E$ we get

\begin{eqnarray}
\label{deltaE}
\Delta E&=&\frac{1}{2}\sum_{k<k_{c}}
\frac{4\pi }{k^{2}} \langle\hat{n}_{\bf k}\rangle_0 
\langle\hat{n}_{-{\bf k}}\rangle_0 \nonumber\\
&=&\frac{1}{2}
\int n^{l}({\bf r}) n^{l}({\bf r}')W({\bf r}-{\bf r}') 
d^{3}r d^{3}r' \;,
\end{eqnarray}

\noindent
where $n^{l}({\bf r})$ is the long wavelength ($k<k_{c}$) part of the
ground-state electron density and $W$ is the periodic (Ewald-summed)
Coulomb interaction.  The constant $\Delta E$ is therefore the long
wavelength contribution to the Hartree energy.  For $\Delta V$ we get

\begin{equation}
\Delta V({\bf r})=-\frac{1}{\sqrt{V}}
\sum_{k<k_{c}}
\frac{4\pi }{k^{2}}
\langle\hat{n}_{\bf k}\rangle_0 e^{-i{\bf k} {\bf r}} \;,
\end{equation}

\noindent
and hence

\begin{equation}
\nabla^{2} \left[\Delta V({\bf r})\right] = 4\pi n^{l}({\bf r}) \;.
\end{equation}

\noindent
$\Delta V({\bf r})$ is therefore the Hartree potential corresponding
to the long wavelength Fourier components of the electronic charge
density.  Because $\Delta V({\bf r})$ is subtracted from $V({\bf r})$
to give $\tilde{V}({\bf r})$, the extended Hamiltonian $\hat{H}_{\rm
BP}$ contains a reduced external potential.  The long-range part of
the mutual repulsion of the electrons has been absorbed into the
plasmon degrees of freedom via the subsidiary condition.

\subsection{The unitary transformation}
\label{alg.2}

We have now concluded that we can concentrate on the ground state of
the Bohm-Pines Hamiltonian, $\hat{H}_{\rm BP}$, from
Eq.~(\ref{eq:hbp}) instead of the ground state of the original
Hamiltonian from Eq.~(\ref{horig}), provided we choose the constants
$\pi^{0}_{\bf k}$ and hence the effective potential $\tilde{V}$ in
accordance with Eq.~(\ref{numb1}).  The oscillator-space part of the
ground state $|\Phi_{0}\rangle$ of $\hat{H}_{\rm BP}$ is then equal to
$|{\mbox{\boldmath $\pi$}^{0}}\rangle$, and the real-space part is the
physical ground state $|\psi_{0}\rangle$.

Eq.~(\ref{numb1}) specifies $\pi^{0}_{\bf k}$ in terms of the
ground-state density, which is obtained by solving $\hat{H}_{\rm BP}$.
Unfortunately, this Hamiltonian depends on the parameters
$\pi^{0}_{\bf k}$ calculated from its ground state, and so we are
faced with a self-consistency problem analogous to those encountered
in bandstructure calculations.  We could in principle devise an
iterative algorithm to home in on a self-consistent solution, but this
is unnecessary.  We only need the ground-state density, not the wave
function itself, and it is known that the LDA gives reasonably good
ground-state densities in most solids.  In practice, therefore, we can
use the LDA density $\langle
\hat{n}_{\bf k}\rangle^{\rm LDA}_0$ to obtain a good approximation for
$\pi^{0}_{\bf k}$:

\begin{equation}
\pi^{0}_{\bf k} \approx \left(\frac{4\pi }{k^{2}}\right)^{1/2}
\langle \hat{n}_{\bf k} \rangle^{\rm LDA}_0 \;.
\end{equation}

Unsurprisingly, the Bohm-Pines extended Hamiltonian cannot be solved
exactly.  It may, however, be solved approximately by means of a
unitary transformation.\cite{Bohm:RPA} We use the unitary operator

\begin{equation}
\label{utr}
\hat{S}=\exp\left[-i\sum_{k<k_c}\left(
\frac{4\pi }{k^{2}}\right)^{1/2}\hat{q}_{\bf k}\hat{n}_{\bf k}\right]
\end{equation}

\noindent
to transform an eigenstate $|\Phi\rangle$ of $\hat{H}_{\rm BP}$ into
an eigenstate

\begin{equation}
|\Phi^{\rm new}\rangle = \hat{S}|\Phi\rangle
\end{equation}

\noindent
of the transformed Hamiltonian $\hat{H}^{\rm new}_{\rm BP} =
\hat{S}\hat{H}_{\rm BP}\hat{S}^{\dagger}$.  (In general, all operators
transform according to $\hat{O}^{\rm new} = \hat{S} \hat{O}
\hat{S}^{\dagger}$.)  The position operators $\hat{\bf r}_{i}$ and
$\hat{q}_{\bf k}$ are unchanged by the transformation because they
commute with $\hat{S}$.  The momentum operators transform as follows:

\begin{eqnarray}
\hat{\bf p}_{i} & \rightarrow & \hat{\bf p}_{i}^{\rm new}
= \hat{\bf p}_{i} +
i\left(\frac{4\pi}{V} \right)^{1/2}
\sum_{k<k_{c}}\hat{q}_{\bf k}
{\mbox{\boldmath $\varepsilon$}}_{\bf k}
e^{i{\bf k} \cdot 
\hat{\bf r}_{i}} \;, \\
\hat{\pi}_{\bf k} & \rightarrow & \hat{\pi}_{\bf k}^{\rm new} 
=  \hat{\pi}_{\bf k} + \left(\frac{4\pi }{k^{2}}\right)^{1/2}
\hat{n}_{\bf k} \;,
\end{eqnarray}

\noindent
where ${\mbox{\boldmath $\varepsilon$}}_{\bf k} = {\bf k}/{k}$ is a
unit vector in the ${\bf k}$ direction. 
The transformations of the momentum operators may be checked using the
general expansion,

\begin{equation}
e^{\hat{X}}\hat{O}e^{-\hat{X}}=\hat{O}+\left[\hat{X},\hat{O}\right]
+\frac{1}{2}\left[\hat{X},\left[\hat{X},\hat{O}\right]\right] + \ldots
\;,
\end{equation}

\noindent
with $\hat{S} = e^{\hat{X}}$.  When $\hat{O}$ is one of the momentum
operators, $\hat{\pi}_{\bf k}$ or $\hat{\bf p}_{i}$, the first
commutator $[\hat{X},\hat{O}]$ only contains position operators, and
so higher order commutators such as $[\hat{X},[\hat{X},\hat{O}]]$ all
vanish.

The final result of the unitary transformation defined by 
Eq.~(\ref{utr}) is the Hamiltonian:

\begin{eqnarray}
\label{brpa}
\lefteqn{
\hat{H}_{\rm BP}^{\rm new} = \frac{1}{2} \sum_{i} \hat{\bf p}_{i}^{2} 
+ 2\pi\sum_{k>k_{c}}
\frac{\hat{n}_{\bf k}^{}\hat{n}^{\dagger}_{\bf k}}{k^2} } &&
\nonumber \\
& - & \frac{2\pi N}{V}\sum_{\bf k} \frac{1}{k^{2}} 
+ \sum_{i} \tilde{V}( \hat{\bf r}_{i})  \nonumber 
\\
& + & i\left(\frac{4\pi}{V}\right)^{1/2}\sum_{k<k_{c}}\sum_{i}
{\mbox{\boldmath $\varepsilon$}}_{\bf k}
\cdot \left( \hat{\bf p}_{i}-\frac{\bf k}{2} \right)
\hat{q}_{\bf k}e^{i{\bf k}\cdot \hat{\bf r}_{i}}
\nonumber \\
& + & \frac{2\pi}{\sqrt{V}}
\sum_{k,k'<k_{c}} \left (
{\mbox{\boldmath $\varepsilon$}}_{\bf k}
\cdot
{\mbox{\boldmath $\varepsilon$}}_{{\bf k}'} \right )
\hat{q}_{\bf k}\hat{q}_{-{\bf k}'}
\hat{n}_{{\bf k}-{\bf k}'}
\nonumber \\
& + & \frac{1}{2}\sum_{k<k_{c}}\hat{\pi}_{\bf k}\hat{\pi}_{-{\bf k}} \;,
\end{eqnarray}

\noindent
which is obtained by replacing $\hat{p}_i$ and $\hat{\pi}_{\bf k}$ in
Eq.~(\ref{eq:hbp}) by $\hat{p}_i^{\rm new}$ and $\hat{\pi}_{\bf
k}^{\rm new}$.  The subsidiary condition becomes

\begin{equation}
\label{sc2}
\hat{\Omega}_{\bf k}^{\rm new}|\Phi^{\rm new}\rangle=0 \;\;\;\;\;\;\; 
(k<k_c) \;,
\end{equation}

\noindent
where

\begin{equation}
\label{sc2.1}
\hat{\Omega}_{\bf k}^{\rm new} = \hat{\pi}_{\bf k} + 
\left(\frac{4\pi }{k^{2}}\right)^{1/2}
\left(
\hat{n}_{\bf k}
- \langle\hat{n}_{\bf k}\rangle_0\right) \;.
\end{equation}

We now make the random-phase approximation, which amounts to replacing
the $\hat{n}_{{\bf k}-{\bf k}'}=V^{-1/2}\sum_{i} e^{i({\bf k}-{\bf
k}')\cdot \hat{\bf r}_{i}}$ factor in the fourth line of
Eq.~(\ref{brpa}) by its ground-state expectation value.  In uniform
systems the electronic positions ${\bf r}_i$ are random and so the
phases are also random; the expectation value of $\hat{n}_{{\bf
k}-{\bf k}'}$ is therefore equal to $N \delta_{{\bf k},{\bf k}'} /
\sqrt{V}$.  In inhomogeneous systems we have to evaluate the
expectation value of the (untransformed) operator $\hat{n}_{{\bf
k}-{\bf k}'}$ in the transformed ground state $|\Phi^{\rm
new}\rangle$.  Since the density operator $\hat{n}({\bf
r})=\sum_{i}\delta(\hat{\bf r}_{i}-{\bf r})$ commutes with the unitary
transformation, it follows that

\begin{equation}
\langle \Phi^{\rm new} | \hat{n}({\bf r}) | \Phi^{\rm new} \rangle 
= \langle \Phi | \hat{S}^{\dagger} \hat{n}({\bf r}) \hat{S} | \Phi 
\rangle = \langle \Phi | \hat{n}({\bf r}) | \Phi \rangle \;.
\end{equation}

\noindent
The required expectation value of $\hat{n}_{{\bf k}-{\bf k}'}$ is
therefore equal to the Fourier component 
$\langle\hat{n}_{{\bf k}-{\bf k}'}\rangle_0$ of
the ground-state electron density of the original (untransformed)
Bohm-Pines Hamiltonian from Eq.~(\ref{eq:hbp}).

The term on the third line of Eq.~(\ref{brpa}) may be rewritten as

\begin{equation}
\frac{i}{2}\left ( \frac{4\pi}{V}\right )^{1/2}\sum_{k<k_{c}}
\hat{q}_{\bf k} 
{\mbox{\boldmath $\varepsilon$}}_{\bf k}
\cdot \hat{\bf j}_{\bf k} \;,
\end{equation}

\noindent
where $\hat{\bf j}_{\bf k} = \sum_{i} \{\hat{\bf p}_{i},e^{i{\bf
k}\cdot \hat{\bf r}_{i}}\}$ is the current density operator.
Following Bohm and Pines,\cite{Bohm:RPA} this plasmon-electron
coupling term will be neglected.  To justify this approximation (and
indeed the RPA) we can appeal to the measured physical properties of
interacting electron gases; we know that the plasmons are well defined
when $k<k_c$, and hence that the plasmon-electron coupling terms must
indeed be small.

A more physical discussion of the RPA may be found in
Appendix~\ref{sec:appA}.

\subsection{The RPA ground state}
\label{alg.3}

The two approximations described above decouple the electrons and
plasmons and reduce the transformed Hamiltonian of Eq.~(\ref{brpa}) to
the RPA Hamiltonian $\hat{H}_{\rm RPA}=\hat{H}_{sr}+\hat{H}_{p}$.  The
first two lines of Eq.~(\ref{brpa}) yield the short-range electronic
Hamiltonian,

\begin{eqnarray}
\label{hsrc}
\hat{H}_{sr} & = & \frac{1}{2} \sum_{i}\hat{\bf p}_{i}^{2} 
+ 2\pi  \sum_{k>k_{c}} 
\frac{\hat{n}_{\bf k} \hat{n}^{\dagger}_{\bf k}}
{k^{2}}
\nonumber \\
& & - \frac{2\pi N}{V}\sum_{\bf k} \frac{1}{k^{2}} 
+ \sum_{i} \tilde{V}( \hat{\bf r}_{i}) \;, 
\end{eqnarray}

\noindent
and the last two lines yield the plasmon Hamiltonian,

\begin{equation}
\label{RPAham}
\hat{H}_{p}=\frac{1}{2}( \hat{\mbox{\boldmath $\pi$}} \cdot
\hat{\mbox{\boldmath $\pi$}}^{\dagger}+ \hat{\bf q}
\cdot M\cdot \hat{\bf q}^{\dagger}) \;,
\end{equation}

\noindent
where the matrix $M$ is given by

\begin{equation}
\label{mkk}
M_{{\bf k}, {\bf k}'}=
({\mbox{\boldmath $\varepsilon$}}_{\bf k}
\cdot
{\mbox{\boldmath $\varepsilon$}}_{{\bf k}'})
\frac{1}{V}
\int e^{i({\bf k}-{\bf k}')\cdot {\bf r}}
\omega^{2}_{p}({\bf r}) d^{3}r \;,
\end{equation}

\noindent
and we have introduced a position dependent \emph{local plasma
frequency} defined by $\omega_p^2({\bf r}) = 4\pi n({\bf r})$.  The
full ground state of $\hat{H}_{\rm RPA}$ is the product of the ground
states of $\hat{H}_{sr}$ and $\hat{H}_{p}$.

\subsubsection{The plasmon ground state}
\label{alg.3a}

If we choose to work in a representation in which the $\hat{\pi}_{\bf
k}$ operators are diagonal, the plasmon ground state takes the
standard simple harmonic oscillator form:

\begin{equation}
\label{PGS}
\Psi_{p}
\propto \exp\left[-\frac{1}{2}\sum_{k,k'<k_{c}} \pi^{\ast}_{\bf k}
\left(M^{-1/2}\right)_{{\bf k},{\bf k}'}
\pi_{{\bf k}'} \right] \;.
\end{equation}

\noindent
The matrix $M^{-1/2}$ is well defined since all the eigenvalues of the
Hermitian matrix $M$ are greater than zero.  The fact that the
$\hat{\pi}_{\bf k}$ and $\hat{q}_{\bf k}$ operators are non-Hermitian
may cause some confusion here, but one can easily rewrite the
$k$-space Hamiltonian of Eq.~(\ref{RPAham}) in real space using
Eqs.~(\ref{opdef1}) and (\ref{opdef2}).  The real-space operators
$\hat{\pi}({\bf r})$ and $\hat{q}({\bf r})$ are Hermitian, and so the
plasmon Hamiltonian is then a set of coupled Hermitian harmonic
oscillators.  The ground state of these oscillators is

\begin{equation}
\Psi_{p}
\propto \exp\left[-\frac{1}{2}\int 
\pi({\bf r})
{M}^{-1/2}({\bf r},{\bf r}') 
\pi({\bf r}')
d^{3}r d^{3}r' 
\right] \;,
\end{equation}

\noindent
which reduces to Eq.~(\ref{PGS}) when re-expressed in $k$-space.

\subsubsection{The short-range ground state}
\label{alg.3b}

If we make use of the expression for $\tilde{V}$ from Eq.~(\ref{npot})
and the condition $\pi^{0}_{\bf k} = \sqrt{4\pi/k^2}
\langle\hat{n}_{\bf k}\rangle_0$ from Eq.~(\ref{numb1}), the
short-range Hamiltonian of Eq.~(\ref{hsrc}) becomes:

\begin{eqnarray}
\label{hsr2}
\lefteqn{ \hat{H}_{sr} =
\frac{1}{2} \sum_{i} \hat{\bf p}_{i}^{2} + 
\sum_i V(\hat{\bf r}_i) } \nonumber \\
& + & 2\pi \sum_{k > k_c}
\frac{\hat{n}_{\bf k}^{ } \hat{n}^{\dagger}_{\bf k}} {k^{2}}
 - \frac{2\pi N}{V} \sum_{\bf k} \frac{1}{k^2} \nonumber \\
& + &
\frac{1}{\sqrt{V}} \sum_i \sum_{k<k_c} 
\frac{4\pi \langle\hat{n}_{\bf k}\rangle_0}{k^2} 
e^{-i {\bf k}\cdot \hat{\bf r}_{i}} \nonumber \\
& - & 2\pi \sum_{k<k_c} 
\frac{\langle\hat{n}_{\bf k}\rangle_0 
\langle\hat{n}_{-{\bf k}}\rangle_0}{k^2}
\;.
\end{eqnarray}

\noindent
The first two lines are identical to the original Hamiltonian,
Eq.~(\ref{horig}), but with the small $k$ (long wavelength)
contributions to the electron-electron interactions omitted.  The
third line is the Hartree potential corresponding to those long
wavelength Coulomb interactions,

\begin{equation}
\frac{1}{\sqrt{V}} \sum_i \sum_{k<k_c} 
\frac{4\pi \langle\hat{n}_{\bf k}\rangle_0}{k^2} 
e^{-i {\bf k}\cdot \hat{\bf r}_{i}}
=
\sum_i \int \frac{n^l({\bf r}')}{|\hat{\bf r}_i - {\bf r}'|} d^3r' \;,
\end{equation}

\noindent
and the fourth line is the Hartree energy, which is subtracted to
prevent double counting.  

The short-range Hamiltonian is therefore equivalent to the original
Hamiltonian, Eq.~(\ref{horig}), but with the long wavelength parts of
the Coulomb interaction treated within the Hartree approximation.
Since the long wavelength parts of the effective potentials used in
Hartree-Fock and LDA calculations are both dominated by the Hartree
contributions, we might equally well say that $\hat{H}_{sr}$ is
equivalent to the original Hamiltonian but with the small $k$ Coulomb
interactions approximated using Hartree-Fock or LDA, provided
the HF or LDA densities are sufficiently similar to
the exact ground state density. The short-range
Hamiltonian still contains the full Coulomb interaction for $k>k_c$,
and so still diverges like $1/r_{ij}$ whenever two electrons approach
each other.  The electron-electron cusps therefore appear in the
short-range electronic wave function, not in the Jastrow factor that
describes the plasmons.

In practice, of course, we do not attempt to solve $\hat{H}_{sr}$
exactly, but treat it within an independent electron approach such as
Hartree-Fock or LDA.  This additional approximation replaces the
short-range part of the electron-electron interaction by a mean field,
which simply adds to the long wavelength mean field already introduced
by the RPA.  The overall effect is equivalent to starting from the
original Hamiltonian and replacing the \emph{full} interaction by a
mean field.  This implies that one can obtain the short-range
``electronic'' part of the RPA wave function by starting from the
original fully interacting Hamiltonian and treating it using any
sensible mean-field approximation.  The best single-particle orbitals
to use in the Slater determinant are therefore very close to the
familiar Hartree-Fock or LDA orbitals; they are not significantly
altered by the presence of the RPA Jastrow factor from
Eq.~(\ref{RPAjas}).

One drawback of treating the short-range Hamiltonian within a
mean-field approximation is that this neglects the electron-electron
cusps that should be present in the short-range electronic wave
function.  The cusps play an important role in reducing the total
energy of the many-electron system, and so the trial wave function may
be significantly improved by building them into the Jastrow factor.

\subsection{Inverting the unitary transformation}
\label{alg.4}

The ground state of $\hat{H}_{\rm RPA}$ is the product of the ground
states of $\hat{H}_{sr}$ and $\hat{H}_{p}$, neither of which commutes
with the transformed subsidiary condition, Eq.~(\ref{sc2.1}).  This
implies that, unlike the ground state of $\hat{H}_{\rm BP}^{\rm new}$,
the ground state of $\hat{H}_{\rm RPA}$ need not obey the subsidiary
condition automatically.  In consequence, the approximate ground state
of $\hat{H}_{\rm BP}$ obtained by applying the back transformation,
$\hat{S}^{\dagger}$, to the ground state of $\hat{H}_{\rm RPA}$, need
not be an eigenfunction of the plasmon momentum operators, and we can
no longer extract an approximation to the spatial ground state by
simply forgetting about the $|{\mbox{\boldmath $\pi$}}\rangle$ factor
in a product wave function of the form $|\psi\rangle|{\mbox{\boldmath
$\pi$}}\rangle$.  Fortunately, however, the subsidiary condition is
still exact (no approximations were made in transforming it), and so
still defines the subspace of the extended Hilbert space in which the
true ground state lies.  We can therefore take the ground state of the
approximate Hamiltonian, $\hat{H}_{\rm RPA}$, and project it onto that
subspace.  The projection operator may be applied before or after the
back transformation, but if we choose to make the back transformation
first it is not difficult to see that the required projection operator
is

\begin{equation}
\label{prop}
\prod_{k<k_{c}}\left|\pi_{\bf k}=\pi^{0}_{\bf k}\rangle
\langle\pi_{\bf k}=\pi^{0}_{\bf k}\right| \;.
\end{equation}

As discussed in Sec.~\ref{alg.3b}, we approximate the ground state of
$\hat{H}_{sr}$ as a Slater determinant $D$, and so the approximate
ground state of the full Hamiltonian $\hat{H}_{\rm RPA}$ is $\Phi_{\rm
RPA} \propto \Psi_{p} D$.  We can now obtain an approximate ground
state of the original Hamiltonian $\hat{H}_{\rm BP}$ by back
transforming using the inverse of the unitary transformation.  The
only important effect of the back transformation is to shift the
numbers ${\pi}_{\bf k}$ appearing in $\Psi_{p}$ by
$-(4\pi/k^{2})^{1/2} n_{\bf k}$:

\begin{equation}
\pi_{\bf k} \rightarrow \pi_{\bf k}
-\left({4\pi }/{k^{2}}\right)^{1/2}n_{\bf k} \;,
\end{equation}

\noindent
where $n_{\bf k}=V^{-1/2}\sum_{i}e^{i{\bf k}\cdot{\bf r}_{i}}$.  This
can be verified by observing that, when evaluating a back-transformed
wave function $\Psi^{\rm old}(\{{\bf r}_{i}\},\{\pi_{\bf k}\})= \langle
\{{\bf r}_{i}\},\{\pi_{\bf k}\}|\hat{S}^{\dagger}|\Psi\rangle$, we can
apply the transformation to the bra $\langle \{{\bf
r}_{i}\},\{\pi_{\bf k}\}|$ rather than the ket $|\Psi\rangle$.  But
since

\begin{eqnarray}
\lefteqn { 
\hat{\pi}_{\bf k} \hat{S} \left | \{ {\bf r}_{i}\},\{\pi_{\bf k}\}
\right \rangle
= \hat{S}  \hat{S}^{\dagger}  \hat{\pi}_{\bf k} \hat{S} 
\left | \{{\bf r}_{i}\},\{\pi_{\bf k}\} \right \rangle }
\nonumber \\
&=& \hat{S}  \left( \hat{\pi}_{\bf k}- 
\left({4\pi}/{k^{2}}\right)^{{1}/{2}}
\hat{n}_{\bf k} \right)
\left | \{{\bf r}_{i}\},\{\pi_{\bf k}\} \right \rangle
\nonumber \\
&=& 
\left( \pi_{\bf k}- \left({4\pi}/{k^{2}} \right)^{{1}/{2}}
n_{\bf k} \right) 
\hat{S} \left | \{{\bf r}_{i}\},\{\pi_{\bf k}\} \right \rangle \;,
\end{eqnarray}

\noindent
we see that

\begin{equation}
\hat{S} \left | \{ {\bf r}_i \}, \{ \pi_{\bf k} \} \right \rangle = 
\left | \{ {\bf r}_i \}, \{ \pi_{\bf k} - 
\left({4\pi}/{k^{2}}\right)^{{1}/{2}}
n_{\bf k} \} \right\rangle \;.
\end{equation}

\noindent
The $\hat{\pi}_{\bf k}$ eigenvalues of the transformed bra are
therefore shifted by $- \left({4\pi }/{k^{2}}\right)^{1/2}n_{\bf k}$
relative to those of the original bra.  As a result, $\Psi^{\rm
old}(\{{\bf r}_{i}\},\{\pi_{\bf k}\}) = \Psi( \{ {\bf r}_{i} \}, \{
\pi_{\bf k} - \left( {4\pi }/{k^{2}} \right)^{1/2} n_{\bf k} \} )$.

Applying the projection operator given in Eq.~(\ref{prop}) replaces
the remaining $\pi_{\bf k}$ by $\pi^{0}_{\bf k} = ({4\pi
}/{k^{2}})^{1/2} \langle\hat{n}_{\bf k}\rangle_0$. (In the homogeneous
case this is zero.)  All in all, then, the spatial part of the
approximation to the ground state is

\begin{equation}
\label{RPApsi}
\Psi\propto\Psi_{J} D
=\exp\left[\frac{1}{2}\sum_{i,j} \tilde{u}({\bf r}_{i},{\bf r}_{j}) 
\right] D \;,
\end{equation}

\noindent
where  

\begin{eqnarray}
\lefteqn { \tilde{u}({\bf r},{\bf r}') = -4\pi \sum_{k,k'<k_{c}} 
\left [ \rule{0cm}{6mm} \right . } 
\nonumber \\
& & 
 \left(\frac{e^{-i{\bf k}\cdot{\bf r}}}{\sqrt{V}}-
\frac{\langle\hat{n}_{-{\bf k}}\rangle_0}{N}\right)
\frac
{\left[M^{-\frac{1}{2}}\right]_{{\bf k},{\bf k}'}}
{k k'} 
\left( \frac{e^{i{\bf k}'\cdot{\bf r}'}}{\sqrt{V}} 
- \frac{\langle\hat{n}_{{\bf k}'}\rangle_0}{N} \right) \left 
. \rule{0cm}{6mm} \right ] .
\nonumber \\
\end{eqnarray}

\noindent

The RPA Jastrow factor includes constant terms, one-electron terms,
and two-electron terms.  The constant terms may be ignored as they
only affect the normalization of the wave function.  The remaining one-
and two-electron terms may then be disentangled and the Jastrow factor
rewritten in the form,

\begin{equation}
\label{RPAjas}
\Psi_{J}
\propto \exp\left[\frac{1}{2}\sum_{i,j} u({\bf r}_{i},{\bf r}_{j})
+\sum_{i}\chi({\bf r}_{i}) 
\right] \;,
\end{equation}

\noindent
where $u({\bf r},{\bf r}')$ and $\chi({\bf r})$ are defined via:

\begin{equation}
\label{two1}
u({\bf r},{\bf r}') = -\frac{1}{V} 
\sum_{k,k'<k_{c}} \hspace*{-2mm}
e^{-i{\bf k}\cdot {\bf r}}
\frac
{4\pi\left[M^{-1/2}\right]_{{\bf k}, {\bf k}'}}
{ k k'} 
e^{i{\bf k}' \cdot {\bf r}'} , 
\end{equation}

\noindent
and

\begin{eqnarray}
\label{fchi1}
\chi({\bf r})&=&
\frac{1}{\sqrt{V}} 
\sum_{k,k'<k_{c}}
e^{-i{\bf k}\cdot {\bf r}}
\frac
{4\pi\left[M^{-1/2}\right]_{{\bf k}, {\bf k}'}}
{k k'} 
\langle\hat{n}_{{\bf k}'}\rangle_0 \\
&=&
-\int_{V} u({\bf r},{\bf r}') n^{l}({\bf r}')d^{3}r' \;,
\end{eqnarray}

\noindent
where, as in Eq.~(\ref{deltaE}), $n^l({\bf r})$ is the long wavelength
$k<k_c$ part of the ground-state electron density.  The derivation of
Eq.~(\ref{fchi1}) made use of the symmetry $M_{{\bf k},{\bf k}'} =
M_{-{\bf k}',-{\bf k}}$, which follows from Eq.~(\ref{mkk}).  In
$k$-space, the relationship between $u$ and $\chi$ takes the form,

\begin{equation}
\label{eq:chiu}
\chi({\bf k}) =
-\sum_{k' < k_c} u({\bf k},{\bf k}') n({\bf k}') \;,
\end{equation}  

\noindent
discussed in Sec.~\ref{subsec:QMC-chi}.

In a homogeneous system, Eq.~(\ref{mkk}) states that $M_{{\bf k},{\bf
k}'} = \omega^{2}_{p} \delta_{{\bf k},{\bf k}'}$.  The $\chi$ function
therefore vanishes and the $u$ function becomes

\begin{equation}
u^{\rm hom}({\bf k},{\bf k}')=
-\frac{4\pi}{\omega_{p}}
\frac{1}{k k'} 
\delta_{{\bf k},{\bf k}'} \;.
\end{equation}

\noindent
Transforming to real space we obtain

\begin{eqnarray}
\lefteqn{ u^{\rm hom}({\bf r},{\bf r}')
\; = \;
u^{\rm hom}(|{\bf r}-{\bf r}'|) }
\nonumber \\
&=&
- \frac{1}{V}\frac{1}{\omega_{p}} 
\sum_{k,k'<k_{c}}
e^{-i{\bf k}\cdot {\bf r}+i {\bf k}'\cdot {\bf r}'}
\frac{4\pi}{k k'} 
\delta_{{\bf k},{\bf k}'}
\nonumber \\
&=&
-\frac{1}{V}\frac{1}{\omega_{p}} 
\sum_{k<k_{c}}
e^{-i{\bf k}\cdot({\bf r}-{\bf r}')}
\frac{4\pi}
{k^{2}} \;.
\end{eqnarray}

\noindent
If $k_c$ is set equal to infinity this gives

\begin{equation}
u^{\rm hom}(|{\bf r}-{\bf r}'|) =
-\frac{1}{\omega_{p}|{\bf r}-{\bf r}'|} \;.
\end{equation}

\noindent
For finite $k_c$, the divergence of $u(|{\bf r}-{\bf r}'|)$ at small
$|{\bf r}-{\bf r}'|$ is suppressed, but the $1/|{\bf r}-{\bf r}'|$
decay at large $|{\bf r}-{\bf r}'|$ remains more or less unaltered.

\subsection{The cusp conditions in inhomogeneous systems}
\label{subsec:inhomcusp}

Section \ref{subsec:QMC-RPAu} explained how cusps may be built in to a
homogeneous RPA Jastrow factor by adding an exponential factor to the
$u$ function:

\begin{equation}
\label{cusp1}
u(r)= -\frac{1}{\omega_{p} r}\left(1-e^{-r/F}\right) \;.
\end{equation}

\noindent
At large $r$ this $u$ function has the $1/(\omega_p r)$ behavior
implied by the RPA, while at small $r$ it tends smoothly towards the
required cusp at $r=0$.  When supplemented by appropriate $\chi$
functions, such Jastrow factors are remarkably successful.  It is
therefore worth considering how we might add cusps to our
inhomogeneous RPA $u$ function.

This is not easy, since the inhomogeneous $u$ function is given as a
complicated truncated double Fourier series.  The series determines
the behavior of $u({\bf r},{\bf r}')$ when ${\bf r}-{\bf r}'$ is
large, and we have to find a way of splicing this known long-range
behavior onto the cusp at small ${\bf r}-{\bf r}'$.  The cusp fixes
the slope of $u({\bf r},{\bf r}')$ as $|{\bf r}-{\bf r}'| \rightarrow
0$, but does not determine its position-dependent value at the point
${\bf r} = {\bf r}'$.  This makes it difficult to implement simple
interpolation schemes that use different functions to describe $u$ at
small and large ${\bf r}-{\bf r}'$.  The introduction of a
multiplicative factor, as in the homogeneous $u$ function of
Eq.~(\ref{cusp1}), is equally problematic.

It turns out that this interpolation problem is easiest to handle when
expressed in $k$-space.  This might seem unlikely at first, since a
true cusp can only be generated by a computationally intractable
infinite Fourier sum.  In practice, however, it appears that
satisfactory approximate cusps can be introduced using Fourier sums
with manageable numbers of terms.  The $k$-space cutoff needs to be
large enough to ensure that the real space volume within which the
cusp is not represented correctly is small, and hence that the
resulting errors contribute little or nothing to expectation values.

Eq.~(\ref{cusp1}) can be Fourier analyzed to give:

\begin{equation}
u({\bf k})=\frac{-4\pi}{\sqrt{V}\omega_{p}}
\frac{{1}/{F^{2}}}{k^2 (k^{2}+{1}/{F^{2}})} \;.
\end{equation}

\noindent
Eqs.~(\ref{eq:f1}) and (\ref{eq:f2}) show that ${1}/{F^{2}}= C
\omega_{p}$, where $C=1$ for antiparallel spins and $C = 1/2$ for
parallel spins.  Hence

\begin{equation}
\label{cusp2}
u({\bf k})=\frac{-4\pi}{\sqrt{V}}
\frac{C}{k^{2}(k^{2}+C \omega_{p})} \;.
\end{equation} 

\noindent
In section \ref{subsec:results-homog} we test
a homogeneous $u$ function defined using a truncated
Fourier series of this form, and find that most of the cusp energy can
be retrieved using a reasonably low cutoff.  

Eq.~(\ref{cusp2}) defines a natural $k$-space crossover, $k_x$, given
by $k_{x}^{2}=C\omega_{p}$.  The terms with $k<k_x$ produce the RPA
behavior at large $r$, while for $k > k_x$ we have $u({\bf k})
\propto {1}/{k^{4}}$, which generates the cusp.  The density
dependence of $k_x$ differs from that of the plasmon cutoff $k_c$ from
Eq.~(\ref{cutoffk}).  It turns out, however, that for typical metallic
densities $k_c$ and $k_x$ are both of the order of $k_F$.  The $k^2 +
C\omega_p$ factor in the denominator of Eq.~(\ref{cusp2}) therefore
allows us to introduce the cusp without significantly affecting the
large $r$ ($k < k_c$) behavior implied by the RPA.  If the density is
extremely small, $k_x$ ($\propto \omega_{p}^{1/2} \propto
n^{{1}/{4}}$) is smaller than $k_c$ ($\propto n^{{1}/{6}}$), and so
the $k$-space method of imposing the cusp is no longer consistent with
the RPA limit we expect when $k < k_c$.

Equation (\ref{cusp2}) suggests a simple $k$-space prescription for
building a cusp into the inhomogeneous Jastrow factor.  We write the
inhomogeneous Jastrow factor as a double Fourier series,

\begin{equation}
u({\bf r},{\bf r}') = \frac{1}{V} \sum_{{\bf k},{\bf k}'} e^{-i{\bf
k}\cdot{\bf r}} u({\bf k},{\bf k}') e^{i{\bf k}'\cdot{\bf r}'} \;,
\end{equation}

\noindent
noting that in a homogeneous system we have $u({\bf k},{\bf k}') =
\sqrt{V} u({\bf k}) \delta_{{\bf k},{\bf k}'}$.  We use this
relationship to rewrite the homogeneous $u$ function of
Eq.~(\ref{cusp2}) in a form suitable for generalization to the
inhomogeneous case,

\begin{equation}
\label{cusp3}
u({\bf k},{\bf k}')=
-\frac{4\pi C}{k k'}
\left( k k' \delta_{{\bf k},{\bf k}'} + 
C \omega_{p} \delta_{{\bf k},{\bf k}'}\right)^{-1} \;,
\end{equation} 

\noindent
interpreting the inversion as that of a (diagonal) matrix.  In the
absence of cusps, we have seen that the homogeneous Jastrow factor may
be obtained from the inhomogeneous one by replacing

\begin{equation}
M_{{\bf k},{\bf k}'}=\frac{1}{V}
\frac{({\bf k}\cdot{\bf k}')}{k k'}\int 
e^{i({\bf k}-{\bf k}') \cdot {\bf r}}
\omega_p^2({\bf r}) d^{3}r
\end{equation}

\noindent
by $\omega^{2}_{p} \delta_{{\bf k},{\bf k}'}$.  As we now wish to
extrapolate from a homogeneous Jastrow factor to an inhomogeneous one,
we do the opposite and replace $\omega_{p} \delta_{{\bf k},{\bf k}'} =
( \omega^{2}_{p} \delta_{{\bf k},{\bf k}'} )^{{1}/{2}}$ by the matrix
square root $M^{{1}/{2}}_{{\bf k},{\bf k}'}$.  Eq.~(\ref{cusp3}) then
becomes

\begin{eqnarray}
u({\bf k},{\bf k}')
& = & -
\frac{4\pi C}{ k k' } 
\left(k k' \delta_{{\bf k},{\bf k}'}+ C M^{1/2}_{{\bf k},
{\bf k}'}\right)^{-1} \nonumber \\
& = & -
\frac{4\pi C}{ ( k k')^{2}} 
\left( \delta_{{\bf k},{\bf k}'}+\frac{C M^{1/2}_{{\bf k},
{\bf k}'}}{k k'}\right)^{-1} \;.
\label{cuspyu}
\end{eqnarray}

\noindent
The matrix to be inverted is no longer diagonal, but remains Hermitian
and positive definite.

If we make the reasonable assumption that the Fourier series for
$\omega_p^2({\bf r})$ converges fairly rapidly, the elements of the
matrix $M$ are constant along the diagonal and fall off as we move
away from the diagonal.  For large $k$ and $k'$, this guarantees that
$u({\bf k},{\bf k}')$ is dominated by the $1/(k k')^2 \approx 1/k^{4}$
prefactor, generating a cusp.  For small $k$ and $k'$ we have

\begin{equation}
u({\bf k},{\bf k}') \approx -\frac{4\pi C}{ (k k')^{2} }
\left(\frac{C M^{1/2}_{{\bf k},
{\bf k}'}}{k k'}\right)^{-1}
=-4\pi
\frac{M^{-1/2}_{{\bf k},
{\bf k}'}}{k k'} \;,
\end{equation}

\noindent
which is the RPA result. 

The $u$ function of Eq.~(\ref{cuspyu}) therefore interpolates smoothly
between the anisotropic long-range correlation term derived from the
inhomogeneous RPA and the cusp at short range. In $k$-space we now
have a continuous crossover from collective behavior to two-particle
scattering instead of a sudden and rather unphysical cutoff at $k_c$.

\subsection{The one-body term}

The introduction of the cusp modifies the $k<k_c$ Fourier components of
the RPA $u$ function and introduces nonzero Fourier components with
$k>k_c$.  In addition, it makes the $u$ function spin dependent,
suggesting that we need a spin-dependent one-body term.  We therefore
generalize our expression for $\chi({\bf k})$ from Eq.~(\ref{eq:chiu})
by extending the wave vector sum to include components with $k>k_c$ and
introducing a sum over spin indices,

\begin{equation}
\label{fchispin1}
\chi_{\uparrow}({\bf k})=
- \sum_{{\bf k}'} \left [
u_{\uparrow\uparrow}({\bf k},{\bf k}')n_{\uparrow}({\bf k}')
+
u_{\uparrow\downarrow}({\bf k},{\bf k}')n_{\downarrow}({\bf k}')
\right ] \;,
\end{equation}

\noindent
with an equivalent formula for $\chi_{\downarrow}({\bf k})$.  In a
spin-unpolarized system, where $n_{\uparrow}({\bf k}') =
n_{\downarrow}({\bf k}') = \frac{1}{2}n({\bf k}')$, this reduces to

\begin{equation}
\label{fchispin2}
\chi_{\uparrow}({\bf k})=
-\sum_{{\bf k}'} 
\frac{1}{2}\left[
u_{\uparrow\uparrow}({\bf k},{\bf k}')
+
u_{\uparrow\downarrow}({\bf k},{\bf k}')
\right]
n({\bf k}') \;.
\end{equation}

\noindent
In the case of a homogeneous correlation term $u$ this further reduces
to

\begin{equation}
\label{fahychispin}
\chi_{\uparrow}({\bf k})=-
\frac{1}{2}\sqrt{V}\left[
u_{\uparrow\uparrow}({\bf k})
+
u_{\uparrow\downarrow}({\bf k})
\right]
n({\bf k}) \;,
\end{equation}

\noindent
as first proposed by Malatesta {\it et al}.\cite{malatesta_1997}

\section{Results}
\label{sec:results}

This section assesses the effectiveness and accuracy of QMC trial wave
functions containing RPA Jastrow factors.
Section~\ref{subsec:results-geom} describes the systems studied and
explains how the results are presented;
Sec.~\ref{subsec:results-homog} considers homogeneous systems; and
Sec.~\ref{subsec:results-inhomog} looks at inhomogeneous systems.

All the results were obtained using trial wave functions of the
standard Slater-Jastrow form, where the spin-up and spin-down Slater
determinants were constructed using accurate LDA orbitals.  The
Jastrow factor contained two- and one-body terms, $u({\bf r}_{i},{\bf
r}_{j})$ and $\chi({\bf r}_{i})$, of various different types.  Note
that from now on we drop the $u({\bf r}_i,{\bf r}_i)$ self-interaction
terms from the Jastrow factor.  This is equivalent to altering
$\chi({\bf r}_i)$ by a negligible amount (ca.\ 5\%).

The energy averages discussed in the rest of this section were all
accumulated using samples of 10000 statistically independent
configurations of all the electrons; the quoted errors are standard
deviations of the mean of the local energy.

\subsection{System geometry}
\label{subsec:results-geom}

In one-electron bandstructure calculations the energy eigenfunctions
$\psi_{\bf k}({\bf r})$ may be obtained by solving the one-electron
Schr\"{o}dinger equation within a single unit cell subject to Bloch
boundary conditions.  Properties of the infinite crystal may then be
obtained by evaluating Brillouin zone integrals (i.e.\ averaging over
boundary conditions).  In QMC calculations, however, we have to take
account of all the electrons simultaneously, and it is no longer
possible to reduce the problem to a single unit cell.  Instead, we
consider a model solid consisting of a finite simulation subject to
periodic (not Bloch) boundary conditions.  The simulation cell is made
as large as possible to minimize the finite-size errors, and normally
consists of several primitive unit cells containing a few hundred
electrons in total.

The simulation cell we choose is the Wigner-Seitz cell of a
face-centered cubic (FCC) lattice.  Since we use periodic boundary
conditions, any electron that moves out through one face of this cell
immediately re-enters through the opposite face.  The simulation-cell
Hamiltonian obeys the same periodic boundary conditions, and hence a
periodic model potential energy is required; we use the potential
energy per cell of an infinite lattice of identical copies of the
simulation cell.  Since the Wigner-Seitz cell of an FCC lattice is
close to spherical, the interactions between electrons in neighboring
copies of the simulation cell are smaller than for most other
geometries, which helps to reduce the finite-size errors. It is
important not to confuse the simulation-cell lattice with the actual
lattice structure of the solid; the simulation cell may contain many
primitive unit cells, and these need not be face-centred cubic.  The
lattice vectors of the FCC simulation-cell lattice will be denoted by
${\bf A}_1$, ${\bf A}_2$, and ${\bf A}_3$, and the corresponding
body-centered cubic (BCC) reciprocal lattice vectors by ${\bf B}_1$,
${\bf B}_2$, and ${\bf B}_3$.

For reasons of computational efficiency, we have chosen to study
electron gas systems subject to external potentials that vary along
the ${\bf B}_3$ direction only,

\begin{equation}
V_{\rm ext}({\bf r})= V_{0} \cos({\bf B}_3 \cdot {\bf r})
\;,
\end{equation}  

\noindent
where $V_0 = 1$ in atomic units.  Since ${\bf B}_3$ is a reciprocal
lattice vector, this choice ensures that the potential has the same
periodicity as the simulation cell.  The electron density and $\chi$
functions, which also vary only in the ${\bf B}_3$ direction, share
this periodicity.

The one- and two-body terms must also respect the periodic boundary
conditions applied to the simulation cell.  This implies that analytic
Jastrow factors based on the $u$ function of Eq.~(\ref{cusp1}) must be
made periodic by including contributions from all the electrons in a
periodic lattice of identical copies of the simulation cell.  Since
the analytic $u$ function decays like $1/r$ at large $r$, the sum of
contributions is evaluated using Ewald summation techniques.
Numerical Jastrow factors calculated from the inhomogeneous RPA are
periodic by construction.  

\begin{figure}[h]
\begin{center}
\leavevmode
\epsfxsize=8cm \epsfbox{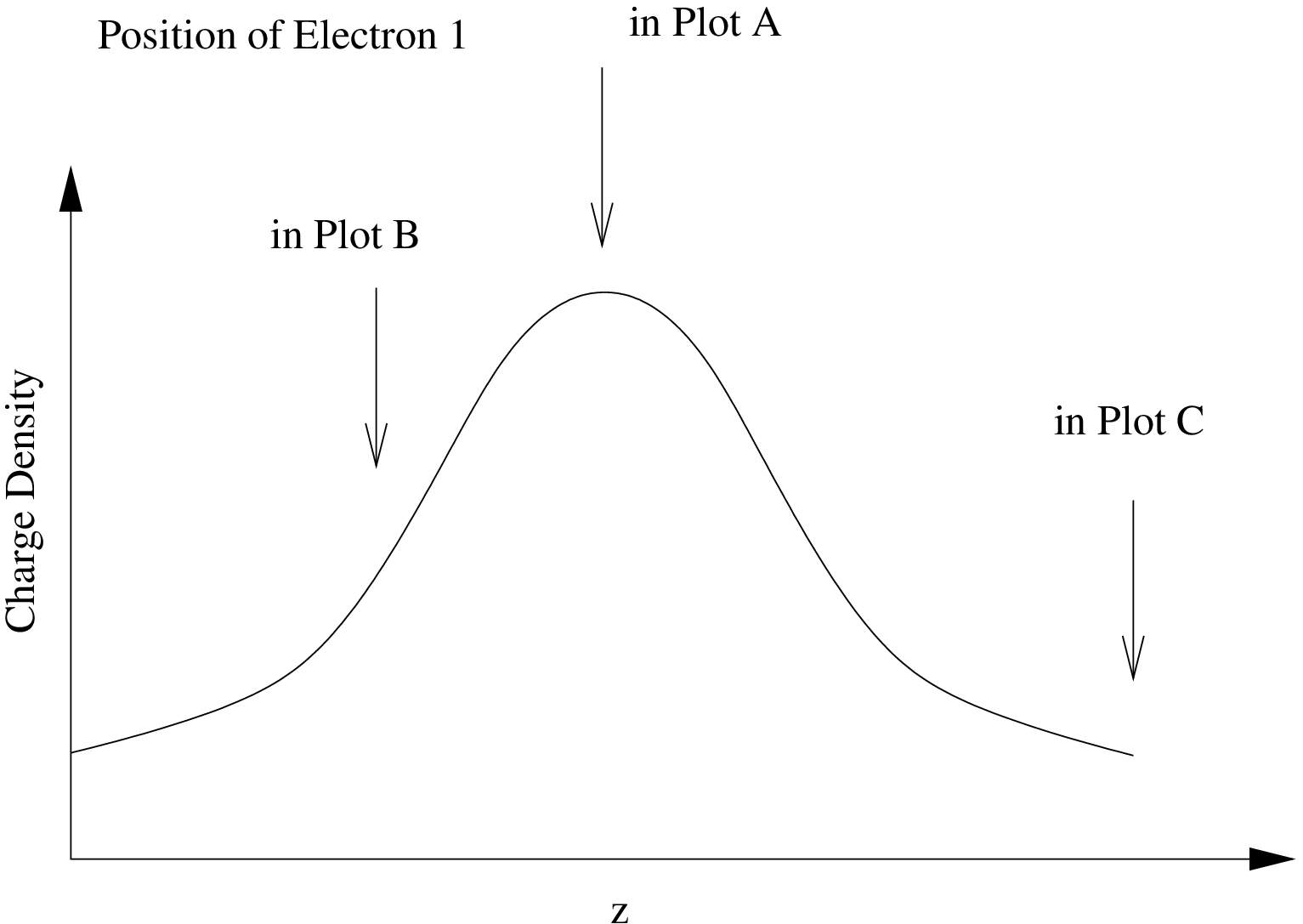}
\end{center}
\caption{ In figures showing two-body terms, plots labeled A show 
$u({\bf r}_1,{\bf r}_2)$ as a function of ${\bf r}_2$ for ${\bf r}_1$
fixed at the peak of the electron density; plots labeled B show
$u({\bf r}_1,{\bf r}_2)$ for ${\bf r}_1$ fixed at the average of the
electron density; and plots labeled C show $u({\bf r}_1,{\bf r}_2)$
for ${\bf r}_1$ fixed at the minimum of the electron density.  In all
cases, ${\bf r}_2$ is swept along the ${\bf B}_3$ direction on a line
passing through ${\bf r}_1$.  The relative coordinate $z$ measures the
distance between the two electrons and thus equals zero when the two
electrons are at the same place, irrespective of the fixed position of
${\bf r}_1$. }
\label{fig_1}
\end{figure}

The next two subsections contain a number of figures showing electron
densities, $\chi$ functions, and $u$ functions.  Charge densities and
$\chi$ functions are plotted along the ${\bf B}_3$ direction from one
side of the simulation cell to the other.  The inhomogeneous two-body
term $u({\bf r}_1,{\bf r}_2)$ is more difficult to represent.  We have
chosen to fix the position ${\bf r}_1$ of the first electron, while
sweeping ${\bf r}_2$ along the ${\bf B}_3$ direction on a line passing
through ${\bf r}_1$.  Figure \ref{fig_1} shows the three positions of
the first electron considered.  Note that because ${\bf A}_3$ and
${\bf B}_3$ are not the same, moving along a line parallel to ${\bf
B}_3$ does not bring you back to a point equivalent to (i.e.\
differing by a lattice vector from) the starting point until you have
passed through three layers of simulation cells.  This means that
although $u({\bf r}_1, {\bf r}_2)$ always has the full periodicity of
the simulation cell, this is not always apparent from the plots.  Note
also that all Jastrow factors are symmetric on interchange of ${\bf
r}_1$ and ${\bf r}_2$.


\subsection{Homogeneous systems}
\label{subsec:results-homog}

The FCC simulation cell considered in this section held a uniform
electron gas of 61 up-spin electrons and 61 down-spin electrons,
giving 122 electrons in total.  The density parameter $r_s$ was equal
to 2, corresponding to a Fermi wave vector $k_F$$=$$0.96a_0^{-1}$.
Two different Jastrow factors were considered:
	
\paragraph{Homogeneous RPA without cusps.}

The homogeneous RPA theory suggests using a correlation term of the
form,

\begin{equation}	
\label{sumrpa}
u({\bf r}_{i},{\bf r}_{j})=
\frac{-4\pi}{V \omega_{p}}\sum_{k<k_c} \frac{1}{k^2}
e^{i{\bf k}\cdot({\bf r}_{i}-{\bf r}_{j})} \;.
\end{equation}

\noindent
We saw in Sec.~\ref{subsec:RPAreview} that for typical metallic
densities the cutoff $k_c$ is set to a value 
comparable to the Fermi wave vector $k_F$.

\paragraph{Homogeneous RPA with cusps.}

In Sec.~\ref{subsec:QMC-RPAu} we saw how a cusp may be introduced by
adding an exponential factor to the $k_c$$\rightarrow$$\infty$ limit
of the homogeneous RPA $u$ function,

\begin{equation}	
\label{rpacusp}
u_{\sigma_i\sigma_j}(r_{ij})=-\frac{1}{\omega_p r_{ij}}
(1-e^{-r_{ij}/F_{\sigma_i\sigma_j}}) \;,
\end{equation}

\noindent
where $F_{\sigma_i\sigma_j}$ is chosen appropriately.

Table \ref{table.conv} shows the local energy averages and standard
deviations obtained in VQMC simulations using these two Jastrow
factors.  For comparison, we also show results obtained using a
``Hartree-Fock'' trial function including up- and down-spin Slater
determinants of LDA orbitals (in this case plane waves) but no Jastrow
factor.  The introduction of an RPA Jastrow factor without a cusp
lowers the calculated energy considerably but has little effect on the
standard deviation.  The introduction of the cusp lowers the energy
greatly and also reduces the standard deviation.  It is clear that the
presence of the cusp is vital if accurate total energies are to be
obtained.


The cuspless RPA results shown in Table \ref{table.conv} were obtained
using a value of $k_c$ equal to the Fermi wave vector $k_F$, but we
also investigated the limit as $k_c$ tends to infinity, in which case
$u(r_{ij})$$\rightarrow$$-1/(\omega_p r_{ij})$.  Since we know that
the description of screening in terms of collective plasmon modes is
invalid at short distances, it was no surprise that this limiting form
gave very poor results.  The residual two-electron interactions in the
short-range Hamiltonian lead to a cusp-like behavior in the wave
function at close distances, not a $-1/(\omega_p r_{ij})$ divergence.

\begin{figure}[h]
\begin{center}
\leavevmode
\epsfxsize=8cm \epsfbox{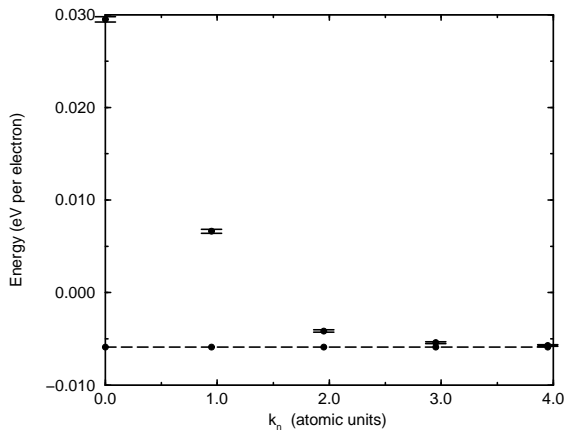}
\end{center}
\caption{ The convergence of the VQMC energy as a function of the 
cutoff $k_n$ used in the truncated Fourier series representation of
the $u$ function from Eq.~(\ref{rpacusp}).  The results are for the
uniform system considered in Sec.~\ref{subsec:results-homog}, for
which $k_{F}=0.96a_0^{-1}$. The dotted line shows the calculated value
of the energy when $k_n$$=$$\infty$ (the standard deviation of this
result is too small to show here).  }
\label{fig_2}
\end{figure}

In Sec.~\ref{subsec:inhomcusp} we saw how the $u$ function with a cusp
from Eq.~(\ref{rpacusp}) may be represented as a Fourier series,

\begin{equation}
\label{kcusp}
u({\bf k})=\frac{-4\pi}{\sqrt{V}}
\frac{C}{k^{2}(k^{2}+C \omega_{p})} \;,
\end{equation} 

\noindent
where $C=1$ for antiparallel spins and $C=1/2$ for parallel spins.  We
can investigate the usefulness of this representation by cutting off
the series at a wave vector $k_{n}$ and varying $k_n$ to see how fast
the calculated VQMC energy approaches the $k_n$$\rightarrow$$\infty$
limit.  The hope is that with a reasonably small $k$-space cutoff we
will be able to produce a Jastrow factor that gives essentially the
$k_n$$\rightarrow$$\infty$ energy.  Figure \ref{fig_2} shows the
convergence of the energy graphically.  It can be seen that a cutoff
of $k_{n}$$=$$3.95$$a_0^{-1}$ produces a wave function with the same
energy (to within statistical uncertainties) as an infinite cutoff.
The $3.95a_0^{-1}$ cutoff is small enough to be computationally
feasible, and so there is no difficulty in representing the cusp in
$k$-space. Note also how the standard deviation tends to zero as the
energy improves.


\subsection{Inhomogeneous systems}
\label{subsec:results-inhomog}

As mentioned above, the inhomogeneous systems we consider have a
background potential that varies in one dimension only.  The LDA
electron density of the unpolarized 64 electron simulation cell
considered in this subsection is shown in Fig.~\ref{fig_3}.  This
system is strongly inhomogeneous (for comparison, a typical
interatomic distance in a solid is $\sim$$6a_0$; this is roughly the
distance between the trough and the peak of the charge density).  The
average electron density is the same as that of a uniform system with
$r_s$$=$$2$ and Fermi wave vector $k_{F}^{0}$$=$$0.96a_0^{-1}$.

In addition to investigating the influence of the cusp, as in the
homogeneous case, we must also now investigate the effects of the
one-body $\chi$ function and compare the accuracies of homogeneous and
inhomogeneous $u$ functions.  We therefore split this section into
three subsections:

\begin{enumerate}

\item[{\em 1.}] First we look at the pure (i.e.\ cuspless) homogeneous
RPA Jastrow factor (which has no $\chi$ function) and compare it with
the pure inhomogeneous RPA Jastrow factor (which does have a $\chi$
function).

\item[{\em 2.}] Second we investigate the effects of adding cusps to
these two correlation factors.

\item[{\em 3.}] Finally we add an ad-hoc one-body $\chi$ term to the
homogeneous RPA Jastrow factor.

\end{enumerate}


\subsubsection{Inhomogeneous RPA without cusps}

The inhomogeneous RPA Jastrow factor considered here is the one
derived in Sec.~\ref{alg.3} (Eqs.~(\ref{RPAjas}), (\ref{two1}), and
(\ref{fchi1})), which includes both $u$ and $\chi$ functions.  As
always in this work, the matrix $M$ is constructed using the LDA
density, and the Slater determinants contain LDA orbitals.  The
homogeneous Jastrow factor includes the $u$ function from
Eq.~(\ref{sumrpa}) but no $\chi$.  In both cases the cutoff $k_c$ is
set equal to the Fermi wave vector, $k_F^{0}$$=$$0.96a_0^{-1}$, of a
homogeneous system with the same average electron density as the
inhomogeneous system.

\begin{figure}[h]
\begin{center}
\leavevmode
\epsfxsize=8cm \epsfbox{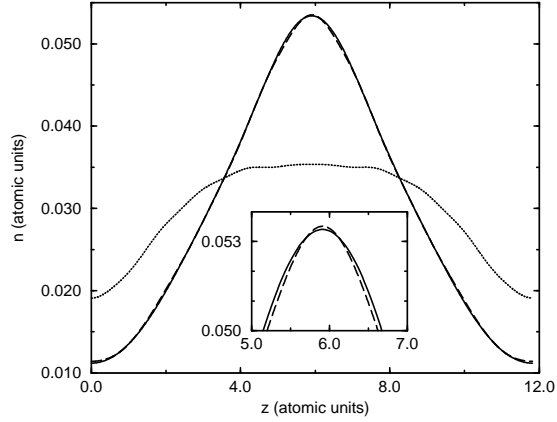}
\end{center}
\caption{ The electron density of the strongly inhomogeneous 64
electron system considered in Sec.~\ref{subsec:results-inhomog}.  The
LDA density (solid line) is compared to the densities obtained using
the homogeneous RPA (dotted line) and the inhomogeneous RPA (dashed
line); the $z$-axis lies along the ${\bf B}_3$ direction.  }
\label{fig_3}
\end{figure}

Table \ref{table2} compares the VQMC energies and standard deviations
calculated using the homogeneous and inhomogeneous Jastrow factors.
It is clear that the inhomogeneous Jastrow factor is much the better
of the two.  The reason is apparent from Fig.~\ref{fig_3}, which
demonstrates that the inhomogeneous Jastrow factor, which has a
built-in one-body term, produces a near optimal density.  Figure
\ref{fig_3} also shows that the homogeneous RPA Jastrow factor (which
has no one-body term) gives a very poor electron density.  This
explains why the corresponding VQMC energy is so poor

\begin{figure}[h]
\begin{center}
\leavevmode
\epsfxsize=8cm \epsfbox{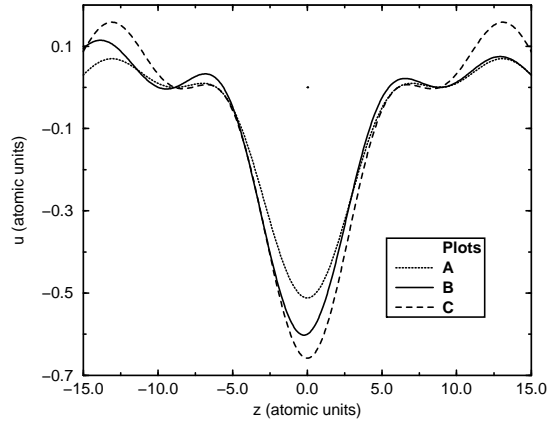}
\end{center}
\caption{ The inhomogeneous RPA $u$ function with no cusp for three 
different positions of the fixed electron.  The results are for the
inhomogeneous system considered in
Sec.~\ref{subsec:results-inhomog}. The definition of $z$ and the
positions of A, B, and C are explained in Fig.~\ref{fig_1}. The
Jastrow factor is stronger in the low density region.  }
\label{fig_4}
\end{figure}

\begin{figure}[h]
\begin{center}
\leavevmode
\epsfxsize=8cm \epsfbox{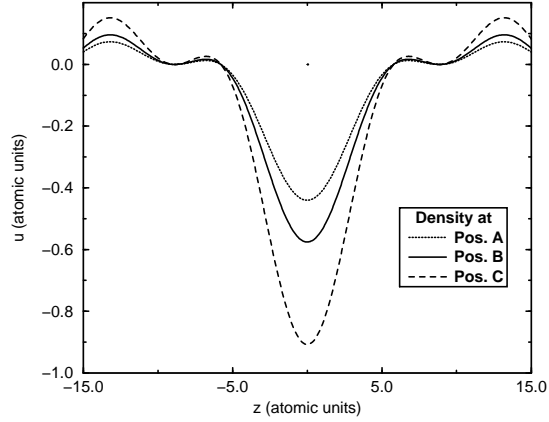}
\end{center}
\caption{ The RPA $u$ functions for three different uniform electron 
gases, the densities of which are equal to the densities at points A,
B, and C of the strongly inhomogeneous 64 electron system considered
in Sec.~\ref{subsec:results-inhomog}. The definition of $z$ and the
positions of A, B, and C are explained in Fig.~\ref{fig_1}. The
homogeneous $u$ functions are of course isotropic. }
\label{fig_5}
\end{figure}

In Fig.~\ref{fig_4} we plot the inhomogeneous RPA two-body term.  Both
inhomogeneity and anisotropy can be seen.  To aid understanding,
Fig.~\ref{fig_5} shows the Jastrow factors of three different
homogeneous systems, the constant densities of which correspond to the
{\it local} densities at the central positions of plots A, B, and C,
respectively.  It is clear that the three inhomogeneous $u$ functions
shown in Fig.~\ref{fig_4} are much more similar than the three
homogeneous $u$ functions shown in Fig.~\ref{fig_5}.  This shows that
the inhomogeneous RPA $u$ function is not well approximated by a
local-density-like approximation based on the homogeneous
RPA.\cite{flad95} In the low density region (position C), in
particular, we find that the inhomogeneous Jastrow factor in
Fig.~\ref{fig_4} is suppressed relative to the ``local density''
version of Fig.~\ref{fig_5}.

At point B, the charge density around the electron is asymmetric, and
this is reflected in the anisotropy of the $u$ function.  The
anisotropy is such that the $u$ function is stronger on the side where
the density is lower; this is consistent with the idea that the RPA
screening is more effective where the electron density is high.

In summary, we find that the Jastrow factor has a larger range in the
low density regions than in the high density regions, consistent with
the ``local density'' picture of Fig.~\ref{fig_4} and with the
physical expectation that screening should be more effective at high
densities.  This interpretation also explains the sign of the
anisotropy: the $u$ function is weaker on the high density side where
the screening is more effective.  We find, however, that the range of
variation of the inhomogeneous Jastrow factor is much smaller than
predicted by the ``local density'' picture.  This suggests that
a homogeneous Jastrow factor defined by the averaged density of the 
underlying inhomogeneous system may not be such a bad approximation.

\subsubsection{Inhomogeneous RPA with cusps}

A cusp may be added to the inhomogeneous RPA $u$ function using the
Fourier-space method explained in Sec.~\ref{subsec:inhomcusp}.  The
results below are obtained with a Fourier cutoff $k_{n}$ of
$4.95a_{0}^{-1}$; Fig.~\ref{fig_2} suggests that this is large enough
to represent the cusp accurately.  Since we choose not to change the
relationship between the $u$ and $\chi$ functions, Eq.~(\ref{fchi}),
the introduction of the cusp also modifies the one-body $\chi$
function.  

The addition of the cusp to the inhomgeneous RPA Jastrow factor
reduces the calculated VQMC energy from $-13.32(4) \times 10^{-2}$ eV
per electron to $-15.81(1) \times 10^{-2}$ eV per electron.  This is
the best variational estimate of the energy we were able to obtain
using any of the Jastrow factors considered in this paper.  The
addition of cusps to the homogeneous $u$ function does not introduce a
one-body term and so the density obtained using the homogeneous RPA
Jastrow factor is still poor.  Energies calculated using the
homogeneous RPA Jastrow factor therefore remain much worse than
energies calculated using the inhomogeneous RPA Jastrow factor.

\begin{figure}[h]
\begin{center}
\leavevmode
\epsfxsize=8cm \epsfbox{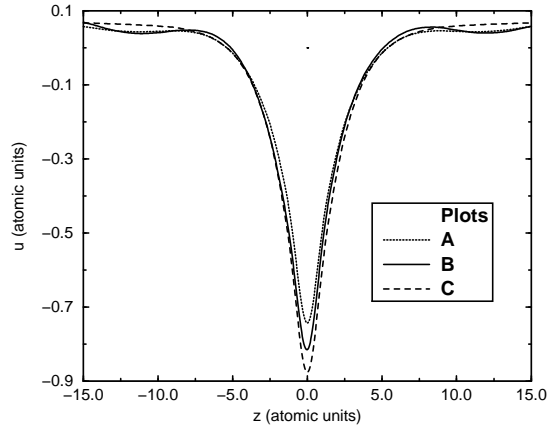}
\end{center}
\caption{ The inhomogeneous RPA $u$ function with a cusp for three 
different positions of the fixed electron.  The results are for the
strongly inhomogeneous 64 electron system considered in
Sec.~\ref{subsec:results-inhomog}.  The definition of $z$ and the
positions of A, B, and C are explained in Fig.~\ref{fig_1}.  The
addition of the cusp has much reduced the inhomogeneity and anisotropy
observed in Fig.~\ref{fig_4}. }
\label{fig_6}
\end{figure}

Figure \ref{fig_6}, which is analogous to Fig.~\ref{fig_4}, shows the
inhomogeneous RPA $u$ function after the cusp has been added.  It is
clear that the addition of the cusp greatly reduces the amount of
inhomogeneity and anisotropy.  Despite the fact that the system is
strongly inhomogeneous, the cusp acts as such a stringent constraint
that $u({\bf r}_i,{\bf r}_j)$ is close to homogeneous.  Although the
inhomogeneity derived from the RPA must persist when $|{\bf r}_i -
{\bf r}_j|$ is large enough, the crossover length, $2\pi/k_x$,
corresponding to the average density $r_s$$=$$2$, is comparable to our
system size.  This implies that the form of the $u$ function is
largely determined by the cusp throughout our system.  If we had
studied larger systems we would have seen the RPA reassert itself at
large $|{\bf r}_i-{\bf r}_j|$, but previous work on numerical trial
function optimization\cite{williamson96} and finite-size
errors\cite{fraser96,williamson97} has shown that the behavior of the
$u$ function at such large values of $|{\bf r}_i - {\bf r}_j|$ has
very little effect on the total energy.  The fact that the
inhomogeneous $u$ function becomes so homogeneous once the cusp has
been added may explain the surprisingly good performance of the
homogeneous $u$ functions used in most QMC simulations of solids.


\subsubsection{Other one-body terms}

In this subsection we compare the quality of analytic one-body terms
based on Eq.~(\ref{fchi}) and numerical one-body terms obtained using
variance optimization.

We have already explained that we always construct the $\chi$ function
appearing in the inhomogeneous RPA Jastrow factor from the $u$
function and density according to Eq.~(\ref{fchi}).  Although
Eq.~(\ref{fchi}) was derived within the RPA, we assume that it holds
unaltered even after the spin-dependent cusps have been added to $u$.
This assumption proves very successful in practice, yielding $\chi$
functions that are not significantly worse than those computed (at
much greater cost) using numerical variance optimization.

\begin{figure}[h]
\begin{center}
\leavevmode
\epsfxsize=8cm \epsfbox{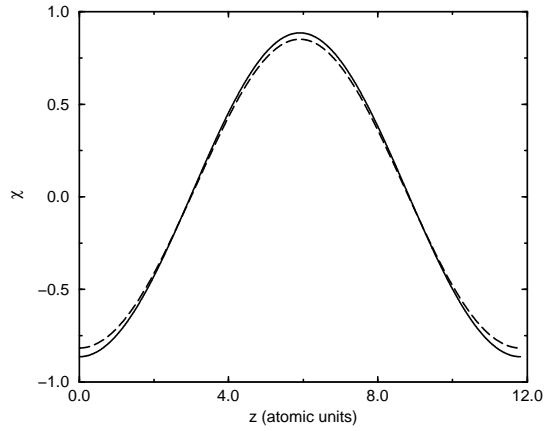}
\end{center}
\caption{ Comparison of the $\chi$ function (solid line) obtained 
from Eq.~(\ref{fahychi}) with one (dashed line) obtained using an
additional variance minimization.  The results are for the strongly
inhomogeneous 64 electron system considered in
Sec.~\ref{subsec:results-inhomog}, using the homogeneous Ewald-summed
Jastrow factor with cusp.  The corresponding energies, shown in Table
\ref{table4}, are equal to within the statistical error. }
\label{fig_7}
\end{figure}

When adding a $\chi$ function to the homogeneous $u$ function of
Eq.~(\ref{rpacusp}) we have two options: we could try using
Eq.~(\ref{fchi}) again, or we could use variance minimization.  Since
$u$ is now a function of $|{\bf r}_i - {\bf r}_j|$ only,
Eq.~(\ref{fchi}) reduces to Eq.~(\ref{fahychi}), which was first
derived by Malatesta {\em et al.}\cite{malatesta_1997,Fahy:CHI} The
impressive accuracy of Eq.~(\ref{fahychi}) is shown in
Fig.~\ref{fig_7}, where we compare the analytic $\chi$ function with
one obtained using additional numerical variance optimization.  We see
that the two $\chi$ functions (and hence the two electron densities)
are very similar.  Table \ref{table4} shows that the difference
between the two VQMC energies is smaller than the statistical noise in
our simulations.



Table \ref{table3} shows that the inhomogeneous RPA (including the
$\chi$ function from Eq.~(\ref{fchi}) and an approximate Fourier
representation of the cusp) yields marginally better results than the
homogeneous RPA (including the $\chi$ function from
Eq.~(\ref{fahychi}) and an approximate Fourier representation of the
cusp).  Comparing these results to the Ewald (i.e.\
$k_{n}$$\rightarrow$$\infty$) results from Table \ref{table4}, we see
that including the exact cusp reduces the variance but has only a
negligible effect on the energy.  The inhomogeneous numerical Jastrow
factor yields a slightly better energy than the Ewald Jastrow factor.
The improvement is barely statistically significant, but suggests that
the inhomogeneity of the RPA correlations does produce a slight
improvement.


\section{Conclusions}
\label{sec:conclusions}

Our aim was to better understand the physics underlying the Jastrow
factors used in QMC simulations of solids, and to derive improved
Jastrow factors for strongly inhomogeneous systems.  We began by
reviewing Bohm and Pines' RPA treatment of the homogeneous electron
gas\cite{Bohm:RPA} and generalizing it to the inhomogeneous case.  The
result of this analysis was a Slater-Jastrow trial wave function
containing an anisotropic inhomogeneous Jastrow factor expressed as a
double Fourier sum.  The optimal orbitals appearing in the Slater
determinants were shown to be close to Hartree-Fock or LDA orbitals,
even though these theories do not include Jastrow factors.

The RPA describes the long-range electronic correlations accurately,
but not the scattering-like correlations at short distances.  We saw,
however, that the correct short-range behavior determined by the cusp
conditions may easily be imposed on any Jastrow factor represented in
$k$-space.  When the inhomogeneous RPA result is modified in this way,
the result is a parameter-free Jastrow factor with the correct short
and long-range behavior.

For systems of a few hundred electrons and an external potential
varying in one dimension only, we showed that trial functions
incorporating modified RPA Jastrow factors are both accurate and
computationally tractable.  Since such Jastrow factors are
parameter-free, the time-consuming variance minimization procedure
normally used to generate accurate $\chi$ functions is not required.
Surprisingly, however, the inhomogeneity of the two-body $u$ function
yields little benefit in the system we studied, producing an energy
only two standard deviations (a barely statistically significant
amount) below the best result obtained using a homogeneous $u$
function.  Provided we have a cusp describing the short-range
interaction and a one-body term to mend the density, the detailed form
of the $u$ function is not very important.  The reason is that the
imposition of the cusp conditions, which fixes the gradient of $u({\bf
r}_i,{\bf r}_j)$ when $|{\bf r}_i-{\bf r}_j| \rightarrow 0$, washes
out most of the inhomogeneity of the RPA $u$ function when $|{\bf
r}_i-{\bf r}_j|$ is small.  It is the short-range correlations that
have most effect on the energy, and so the long-range inhomogeneities
that remain after the cusp conditions have been imposed have little
effect.

If we compare the plasmon cutoff $k_c$ ($\propto n^{1/6}$) from
Eq.~(\ref{cutoffk}) with the wave vector $k_x$ ($\propto n^{1/4}$)
that characterizes the crossover from screening behavior to cusp-like
behavior (see Sec.~\ref{subsec:inhomcusp}), we see that the cusp is
relatively less important in high density systems.  This suggests that
the inhomogeneity of the RPA $u$ function may produce a more obvious
improvement when the average electron density is both large and
strongly varying.  This is intuitively sensible, since in low density
systems we expect the short-range electron-electron scattering
described by the cusp to dominate, whereas at higher densities
screening and collective effects should be more important.  Possible
candidates for high density systems include calculations explicitly
involving the core electrons, where it is already known that the use
of inhomogeneous $u$ functions is advantageous.\cite{umrigar_1988}
Other systems where one might expect inhomogeneities in the
correlation term to become important are rare earth elements, where
some of the valence electrons are strongly bound to the core.

In summary, we hope that this paper has contributed a better and more
general understanding of the physics underlying the Slater-Jastrow
trial functions used in most QMC simulations of solids.

\appendix

\section{Physical interpretation of the random-phase approximation}
\label{sec:appA}

In this appendix we look at the physical interpretation of the
Bohm-Pines Hamiltonian $\hat{H}_{\rm BP}^{\rm new}$ \emph{after} the
application of the unitary transformation discussed in
Sec.~\ref{alg.2}.  As expressed in Eq.~(\ref{brpa}) this Hamiltonian
appears very complicated.  However, if we define a field

\begin{equation}
 \hat{\bf A}({\bf r})=
\left(\frac{4\pi}{V}\right)^{\frac{1}{2}}
\sum_{k<k_{c}}i\hat{q}_{\bf k}{\mbox{\boldmath $\varepsilon$}}_{\bf k}
e^{i{\bf k}\cdot{\bf r}}
\end{equation}

\noindent
and then calculate $\hat{\bf E}({\bf r})= -\frac{d}{dt}{\hat{\bf
A}}({\bf r}) = -i\big[\hat{H}_{\rm BP}^{\rm new},\hat{{\bf A}}({\bf
r})\big]$, we obtain:

\begin{equation}
\hat{\bf E}({\bf r}) =
-\left(\frac{4\pi}{V}\right)^{\frac{1}{2}}
\sum_{k<k_c}i\hat{\pi}_{-{\bf k}}
{\mbox{\boldmath $\varepsilon$}}_{\bf k}
e^{i{\bf k}\cdot{\bf r}} \;.
\end{equation}

\noindent
The transformed Bohm-Pines Hamiltonian may then be written in the much
simpler form:

\begin{eqnarray}
\label{Hfield}
\hat{H}_{\rm BP}^{\rm new} &=&\frac{1}{2}
\sum_{i} \left[ \hat{\bf p}_{i}+ \hat{\bf A}(\hat{\bf r}_{i})\right]^2
-\frac{2\pi N}{V}\sum_{\bf k} \frac{1}{k^{2}}
\nonumber \\
&+&\frac{1}{8\pi}\int_{V} \left[\hat{\bf E}({\bf r})
\right]^{2}d^{3}r
+\sum_{i} \tilde{V}(\hat{\bf r}_{i})  \;.
\end{eqnarray}

\noindent
The Hamiltonian of Eq.~(\ref{Hfield}) describes a set of quantum
mechanical particles moving in a reduced external potential potential
$\tilde{V}({\bf r})$ and interacting with a quantum mechanical
longitudinal electromagnetic field $\hat{\bf A}({\bf r})$.  The
kinetic energy of the field is associated with the energy density of
the electric field in the usual way.  The interactions between the
particles and the field are also described via the standard coupling
to the momentum operators.

The physical interpretation of the transformed subsidiary condition,
$\hat{\Omega}_{\bf k}^{\rm new}|\Phi^{\rm new}\rangle=0$, where
$\hat{\Omega}_{\bf k}^{\rm new}$ is as given in Eq.~(\ref{sc2.1}),
also becomes much clearer when re-expressed in terms of the new
fields; it simply sets the Fourier components of

\begin{equation}
\hat{\Omega}({\bf r}) \; \propto \;
\mbox{div} \hat{\bf E}({\bf r}) -
4\pi\left [ \hat{n}({\bf r})-n({\bf r}) \right ]
\end{equation}

\noindent
to zero.  The subsidiary condition thus ensures that the electric
field is related to the density of the particles via Gauss's law.
Crucially, we found that this constraint is automatically satisfied in
the ground state, provided the ${\bf \pi}^{0}$ have been chosen
correctly.

The RPA decouples the electronic and field variables by neglecting the
$\hat{\bf p}\cdot\hat{\bf A}$ terms and treating the quadratic
$\hat{\bf A}^2$ terms only approximately by averaging the electron
positions.  The field part of the decoupled Hamiltonian is then
harmonic:

\begin{equation}
\label{eq:fieldreal}
\frac{1}{2}\int_{V} \left[\hat{\bf A}({\bf r})
\right]^{2} n({\bf r})d^{3}r
+\frac{1}{8\pi}\int_{V} \left[ \hat{\bf E}({\bf r})
\right]^2 d^{3}r \;.
\end{equation}

\noindent
The resulting wave function no longer obeys Gauss's law automatically,
but a projection operator can be applied to produce a wave function
that does.

Eq.~(\ref{eq:fieldreal}) looks disconcertingly simple: it appears that
we have three independent harmonic oscillators at every point ${\bf
r}$ (one each for the x, y and z components of the vector
potential). The apparent simplicity is misleading, however, since the
condition that the fields be longitudinal, $\mbox{curl} {\bf A} = 0$,
couples the oscillators at different points in space and reduces the
number of degrees of freedom at any point from three to effectively
just one. Dropping this condition and retaining only one scalar
oscillator is equivalent to dropping the ${\mbox{\boldmath
$\varepsilon$}}_{\bf k} \cdot {\mbox{\boldmath $\varepsilon$}}_{{\bf
k}'}$ term in Eq.~(\ref{mkk}) and leads to a local-density-like
version of the RPA.


\begin{table}[h]\centering
\caption{ VQMC local energy averages and standard deviations of the
uniform system considered in Sec.~\ref{subsec:results-homog}.  Results
for three different trial wave functions are shown.  The HF trial
function has no Jastrow factor.  The RPA results use the ``pure'' RPA
$u$ function from Eq.~(\ref{sumrpa}).  The best energies are obtained
using the RPA $u$ function with a cusp from Eq.~(\ref{rpacusp}).  }
\begin{tabular}{ccccc}
\rule{0mm}{4mm}&HF&RPA&RPA+CUSP\\
\hline	
\rule{0mm}{5mm}
Energy~~~($10^{-2}$eV per electron)&2.95&1.94&-0.59&\\
Std.\ dev.\ ($10^{-4}$eV per electron)&2.79&2.60&0.50&\\
\end{tabular}
\label{table.conv}
\end{table}

\begin{table}[h]\centering
\caption{ VQMC local energy averages and standard deviations of the
inhomogeneous system considered in Sec.~\ref{subsec:results-inhomog}.
The HF trial function has no Jastrow factor.  The RPAI trial function
includes a Jastrow factor containing the homogeneous RPA $u$ function
from Eq.~(\ref{sumrpa}) but no $\chi$ function.  The RPAII trial
function uses the full inhomogeneous RPA Jastrow factor derived in
Sec.~\ref{alg.3}.}
\begin{tabular}{ccccc}
\rule{0mm}{4mm}&HF&RPAI&RPAII\\
\hline	
\rule{0mm}{5mm}
Energy~~~($10^{-2}$eV per electron)&-12.24&-5.8&-13.32&\\
Std.\ Dev.\ ($10^{-4}$eV per electron)&4.0&4.4&3.6&\\
\end{tabular}
\label{table2}
\end{table}

\begin{table}[h]\centering
\caption{ VQMC local energy averages and standard deviations of the
system considered in Sec.~\ref{subsec:results-inhomog}. Results for
three different trial wave functions are shown.  The HF trial function
has no Jastrow factor.  The EW trial function includes the
Ewald-summed $u$ function from Eq.~(\ref{rpacusp}) and an analytic
$\chi$ function calculated using Eq.~(\ref{fahychi}).  The VM trial
function uses the same $u$ function but optimizes the Fourier
components of $\chi$ using variance minimization. }
\begin{tabular}{ccccc}
\rule{0mm}{4mm}&HF&EW&VM&\\	
\hline
\rule{0mm}{5mm}
Energy~~~($10^{-2}$eV per electron)&-12.244&-15.786&-15.790&\\
Std.\ Dev.\ ($10^{-5}$eV per electron)&40.0&8.2&8.4&\\
\end{tabular}
\label{table4}
\end{table}

\begin{table}[h]\centering
\caption{ 
VQMC local energy averages and standard deviations of the system
considered in Sec.~\ref{subsec:results-inhomog}. Results for three
different trial wave functions are shown.  The HF trial function has
no Jastrow factor.  The RPAI trial function includes the homogeneous
RPA $u$ function with Fourier components given by Eq.~(\ref{kcusp})
plus a $\chi$ function generated using Eq.~(\ref{fahychi}). The RPAII
trial function uses the inhomogeneous RPA Jastrow factor from
Sec.~\ref{alg.3}, to which cusps have been added as explained in
Sec.~\ref{subsec:inhomcusp}.  The Fourier cutoff $k_n$ was set to
$4.95a_0^{-1}$ in both cases. }
\begin{tabular}{ccccc}
\rule{0mm}{4mm}&HF&RPAI&RPAII\\
\hline	
\rule{0mm}{5mm}
Energy~~~($10^{-2}$eV per electron)&-12.244&-15.785&-15.812&\\
Std.\ Dev.\ ($10^{-5}$eV per electron)&40.0&12.2&11.4&\\
\end{tabular}
\label{table3}
\end{table}

\end{document}